%% file: main.tex
\documentclass[journal]{IEEEtran}

\input{packages}

\ifCLASSINFOpdf
\else
\fi
\begin{document}
%
%
%
\title{Tracing, Ranking and Valuation of Aggregated\\ DER Flexibility in Active Distribution Networks}
%
%
%
\author{Andrey~Churkin,
        Wangwei~Kong,~\IEEEmembership{Member, IEEE},
        Jose~N.~Melchor~Gutierrez,
        Eduardo~A.~Martínez~Ceseña,~\IEEEmembership{Member,~IEEE},
        and~Pierluigi~Mancarella,~\IEEEmembership{Senior Member, IEEE}
\thanks{This work was carried out as a part of the ATTEST project (the Horizon 2020 research and innovation programme, grant agreement No 864298).}
\thanks{The authors are with the Department of Electrical and Electronic Engineering, the University of Manchester, UK (e-mails: \{andrey.churkin, wangwei.kong, jose.melchorgutierrez, p.mancarella, alex.martinezcesena\}@manchester.ac.uk)}
\thanks{P. Mancarella is also with the Department of Electrical and Electronic Engineering, the University of Melbourne, Australia.}
\thanks{E. A. Martinez Cesena is also with the Tyndall Centre for Climate Change Research, UK.}
}

\maketitle
\begin{abstract}
\color{black}
The integration of distributed energy resources (DER) makes active distribution networks (ADNs) natural providers of flexibility services. However, the optimal operation of flexible units in ADNs is highly complex, which poses challenges for distribution system operators (DSOs) in aggregating DER flexibility. For example, to maximise the provision of services, flexible units must be strongly coordinated to manage network constraints, e.g., perform power swaps. Furthermore, due to the nonlinearities of aggregated DER flexibility provision, some units may need to rapidly change their outputs to enable the services.
To address these challenges, this paper brings together exact AC optimal power flow (OPF) models and a cooperative game formulation and presents a new framework for tracing, ranking, and valuation of aggregated DER flexibility in ADNs.
Extensive tests and simulations performed for the 33-bus radial distribution network demonstrate that the framework enables translating complex DER interactions into useful information for DSOs by ranking the criticality of flexible units and performing flexibility valuation based on its cost or economic surplus.
Additionally, the framework proposes no-swap constraints and a nonlinearity metric which can be used by DSOs to identify unreliable operating regions with power swaps or rapid changes in flexible unit dispatch.

\end{abstract}


%
\IEEEpeerreviewmaketitle

\section{Introduction}
%
%
%
%

\IEEEPARstart{T}{he} increasing integration of distributed energy resources (DER) and flexible consumers makes active distribution networks (ADNs) natural providers of flexibility services \cite{Eid2016,Chowdhury2009}. Such services can be utilised within distribution networks, as well as aggregated and traded between distribution system operators (DSOs) and transmission system operators (TSOs). In this regard, multiple TSO-DSO coordination schemes and flexibility market designs have been proposed to enable flexible power trading between distribution and transmission systems \cite{Gerard2018,Vicente-Pastor2019,Schittekatte2020,LeCadre2019,Givisiez2020}. 
More specifically, in recent years, significant research efforts have been devoted to modelling the aggregated flexibility at a selected location (e.g., TSO-DSO interface) as sets of feasible operating points in the P-Q space \cite{Heleno2015,Silva2018,Contreras2018,Capitanescu2018,Riaz2021,Stankovic2020,Kalantar-Neyestanaki2020,Tan2020}. Such sets are known as flexibility P-Q areas, flexible power capability charts, or nodal operating envelopes. 
A thorough comparison of models and approximation methods for aggregated flexibility estimation is given in \cite{Lopez2021,Bolfek2021,Contreras2021}.

Regardless of their benefits, current studies on flexibility P-Q areas focus on estimating the limits (boundary) of the aggregated flexible power provision, overlooking the optimal flexible unit dispatch and contributions of individual units to aggregated flexibility.\footnote{Flexible units are resources with the technical ability to adjust their power exchange with the grid, e.g., controllable DER, such as battery energy storage systems, prosumers, electric vehicle aggregators, etc.} Moreover, the P-Q areas are formulated under the potentially unrealistic assumption that all flexible units are perfectly coordinated {\color{black} and can perform fast flexible power control}. 
{\color{black} However, the complexity and nonlinearity of the optimal flexible unit dispatch can pose significant challenges for DSOs. This work demonstrates two major challenges to the aggregated DER flexibility provision:
\begin{enumerate}
\item \textbf{Flexible power swaps} can occur when flexible units cannot follow purely economic incentives due to network constraints (e.g., voltage and thermal limits). In such cases, some units have to manage network constraints to enable other units to provide flexibility services (and potentially earn most of the revenue from the provision of services). For example, some units may need to produce flexible power (to manage voltage and thermal constraints) whereas other units consume it. DSOs can provide aggregated flexibility in such regimes only under the assumption of perfect unit coordination. However, this assumption may be unrealistic, as some units may only be partially controllable by DSOs, and not all units may exchange information. Recent studies, such as \cite{Petrou2021}, highlighted the need for developing new modelling approaches for distribution networks with flexible resources located behind the meter, e.g., prosumers. Without sufficient control and coordination of flexible units, DSOs would not be able to operate in regimes with flexible power swaps and achieve full aggregated network flexibility. Moreover, using power swaps can be risky for DSOs, as loss of coordination between units can result in infeasible network operation.
\item \textbf{Rapid nonlinear changes in flexible unit dispatch} may be required between close operating points. Such shifts in the flexible power of units are justified by both economic reasons (differences in the units’ cost functions) and network constraints (impact of the binding constraints on unit operation). However, in reality, flexible units might not have sufficient ramp capabilities to follow such a complex dispatch with rapid nonlinear changes. 
Recent projects such as \cite{PowerPotential2021} demonstrated that flexible resources in real networks may have very limited ramp capabilities. It follows that DSOs may face problems in the optimal flexible unit dispatch where slow units may not be able to perform rapid regulation of their power, thus reducing the economic efficiency of the flexibility services or even putting the system operation at risk.
\end{enumerate}

In view of the flexible unit dispatch complexity and the aforementioned challenges, it becomes necessary to develop novel approaches for tracing, ranking and valuation of aggregated DER flexibility in ADNs. Moreover, to accurately capture the nonlinearity and nonconvexity of the aggregated DER flexibility provision, such approaches must include full AC power flow models.
}


There have been attempts to address the problem of flexibility aggregation and disaggregation in ADNs \cite{Muller2019,Yi2021,Fruh2022,Nazir2022,Sarstedt2022}. However, existing literature focuses on either technical (e.g., capacity disaggregation, dispatch feasibility) or economic aspects (e.g., cost minimisation) to assess the contributions of flexible units to aggregated flexibility. Neither of these options is optimal. If flexible units are only ranked based on the maximum flexible power that they can provide (capacity-based ranking), their costs and placement cannot be reflected in flexibility valuation mechanisms. If the ranking is based on cost minimisation, the cheaper flexible units are allocated most of the service payments, whereas the value and contributions by more expensive units, which may perform power swaps to enable the services, are underestimated. Such units may receive fewer incentives to participate in the flexibility service provision. It follows that comprehensive flexibility tracing, ranking, and valuation mechanisms should be able to consider multiple relevant factors, such as the capacities and costs of the flexible units, as well as the nonlinear effects of network constraints. In view of new challenges in DER valuation \cite{PICCIARIELLO2015370}, it becomes necessary to develop adequate valuation mechanisms that provide the right coordination and incentives for flexible units in distribution networks, e.g., availability and delivery (utilisation) payments may be needed to guarantee enough flexibility volume on the market \cite{Ruwaida2022,Sarstedt2022}.

There are studies on independent DER flexibility service modelling that present key concepts and tools for flexibility valuation mechanisms discussed above. For example, existing literature provides a wide range of models to explore DER bidding in local flexibility markets and methods for distribution locational marginal pricing \cite{Pourghaderi2022,Attarha2022,Nair2022,Jian2022,Tsaousoglou2021,Zarabie2019}, as well as DER provision of flexibility and ancillary services in the context of peer-to-peer market design, e.g., in \cite{Zhang2020,Luth2018}. But, as these references focus on distributed solutions for individual DER, they do not explicitly capture the DER coordination and incentives required for aggregated network flexibility studies. Some of the studies also lack the detailed AC models required to analyse the complex interactions between DER in the context of flexibility aggregation.

To capture the value of DER coordination, several studies have explored the cooperative game formulations as a potential solution for specific applications, e.g., prosumer management \cite{Han2019}, energy communities \cite{Hupez2021,Fleischhacker2022,Cremers2023}, and storage sharing systems \cite{Ai2022}.
It has been demonstrated that solution concepts from Cooperative Game Theory, such as the Shapley value, have useful properties for estimating both the economic value of coordination (e.g., costs or payments) and technical aspects (e.g., analysing the effects of constraints and criticality of parameters). Cooperative game formulations have been successfully used for solving allocation and ranking problems in power systems \cite{Churkin2021review,Banez-Chicharro2017,Hasan2018,Kristiansen2018,Azuatalam2019}, analysing the stability and incentive compatibility of cooperation \cite{Fleischhacker2022,Anibal2022,Churkin2022enhancing}, and developing data valuation mechanisms and data marketplaces \cite{agarwal2019marketplace,Goncalves2021}. Yet, no framework has been proposed to use cooperative game formulations for tracing, ranking, and valuating aggregated flexibility in ADNs.

In the context of aggregated DER flexibility valuation, the literature does not consider game-theoretical approaches to capture the value of cooperation and propose coordination mechanisms. Studies such as \cite{Sarstedt2022} considered only cost-minimising disaggregation of flexibility without a comprehensive flexibility ranking and valuation, overlooking contributions of some critical units (such as the ones engaged in power swaps to enable the services). Regardless of this gap, only one study that considers cooperative games for the valuation of aggregated DER flexibility was identified by the authors \cite{Anibal2022}. However, this study used game-theoretical approaches to allocate the cost of flexibility to different system operators rather than to allocate the revenues among DER. Moreover, a linear DC power flow model was utilised, which overlooks the physics of flexible power provision in ADNs. Thus, this approach is not suitable for the valuation of aggregated DER flexibility within ADNs.

To address the aforementioned gaps, this paper proposes a framework for tracing, ranking, and valuating aggregated flexible power in ADNs with multiple flexible units. 
Using an exact AC optimal power flow (OPF) formulation for radial distribution networks and the concept of flexibility P-Q areas, the framework introduces models for estimating the limits of aggregated network flexibility and minimising the cost of flexibility services, therefore providing the optimal flexible unit dispatch. For each feasible flexibility request, these models identify the contributions of flexible units to the capacity, cost, and economic surplus of aggregated flexibility provision.
To deal with the combinatorial nature of flexibility aggregation from multiple flexible units, a cooperative game formulation is introduced. The useful properties of the Shapley value are leveraged to estimate the criticality of flexible units to the aggregated ADN flexible capacity (capacity-based ranking) and allocate payments between units for flexibility provision (surplus-based remuneration).
Even though AC power flow models, the concept of flexibility P-Q areas, and game-theoretical approaches have been used individually in the literature on aggregated network flexibility, the proposed framework is the first one to bring these models together to introduce the dedicated mechanisms for tracing, ranking and valuation of DER flexibility in ADNs.
Moreover, in contrast to existing studies, the framework enables analysing the nonlinearities of the aggregated flexibility provision and potential issues of unit coordination. {\color{black}Binary variables corresponding to P-Q flexibility regulations and no-swap constraints are introduced to identify flexible power swaps between units. A nonlinearity metric and a nonlinearity assessment algorithm are proposed to detect rapid nonlinear changes in flexible unit dispatch.}

{\color{black}
The proposed tracing, ranking, and valuation tools based on the Shapley value facilitate translating complex DER interactions into more useful information for DSOs. Conventional OPF models consider a single snapshot of the aggregated flexibility provision for a given P-Q operating point (only the optimal combination of flexible units). Given the complexity of the optimal flexible unit dispatch, such models can underestimate contributions by some critical flexible units, e.g., units that perform flexible power swaps to alleviate network constraints. In this regard, the proposed game-theoretic formulation captures both the individual contributions of flexible units to flexibility requests and all possible combinatorial effects of joint flexibility provision, providing a comprehensive ranking of the units’ value.}
The applicability of the proposed framework is demonstrated through extensive simulations for a well-known 33-bus radial test system.

The research gap in the literature and the contributions of this work are further highlighted in the Appendix, which provides a mapping of the most relevant references and research directions.
Specifically, the paper makes the following contributions:
\begin{itemize}
  \item A novel framework is introduced for tracing, ranking, and valuating aggregated flexible power in distribution networks. 
  The framework brings together full AC OPF models for radial networks and a cooperative game formulation, which enables estimating contributions of flexible units to aggregated network flexibility and to each separate flexibility request. It can be used by DSOs to identify the most critical flexible units or remunerate units for participating in the flexibility services provision.
  \item It is demonstrated that flexible units exhibit complex nonlinear behaviour when providing aggregated flexibility in ADN. Some units can be required to shift their power output rapidly to perform active network management and voltage control. A nonlinearity metric is proposed to identify operating regions with rapid nonlinear changes in the flexible unit dispatch, which DSOs may want to avoid. The flexible power swap phenomenon is discovered, which happens when different units simultaneously produce and consume flexible power to alleviate network constraints and maximise network flexibility. This behaviour poses challenges for both the operation and pricing of flexible units.
  \item It is shown that, in the context of aggregated flexibility valuation, the Shapley value can be used as a mechanism that inherently includes availability and delivery (utilisation) payments for flexible units. The inclusion of both the optimal unit dispatch and units' potential contributions in the remuneration mechanism can give additional incentives for units to declare their maximum capability at a lower cost.
  
\end{itemize}

The rest of this paper is organised as follows. Section~\ref{Section: models} introduces models and metrics used to characterise aggregated flexibility of distribution networks, formulates the framework for coalitional analysis of flexibility requests, and defines tools for analysing unit coordination and the nonlinearity of aggregated flexible power provision. In Section~\ref{Section: results}, extensive simulations are performed 
for a radial distribution network with four flexible units. The advantages and challenges of the proposed framework are discussed in Section~\ref{Section: discussion}. Finally, Section~\ref{Section: conclusion} concludes the paper. The Appendix provides a mapping of the most relevant references and research directions.

\section{Modelling Framework: Approaches and Metrics to Characterise Network Flexibility}\label{Section: models}

\subsection{Network Flexibility as a Set of Feasible Operating Points}
Assuming the impacts of the distribution network on aggregated flexibility are negligible, the operation limits of multiple flexible units can be mathematically described using the Minkowski addition of their P-Q capability sets \cite{Riaz2021,AGARWAL2002}. For example, considering two sets $A$ and $B$, the Minkowski addition is the set formed by adding each vector in $A$ to each vector in $B$, as denoted by:
\begin{IEEEeqnarray}{l}
    \label{Eq: Minkowski}
    A \oplus B = \{a + b \enskip | \enskip a \in A, b \in B \}
\end{IEEEeqnarray} 

\begin{figure}[b]
    \centering
    \includegraphics[width=\columnwidth]{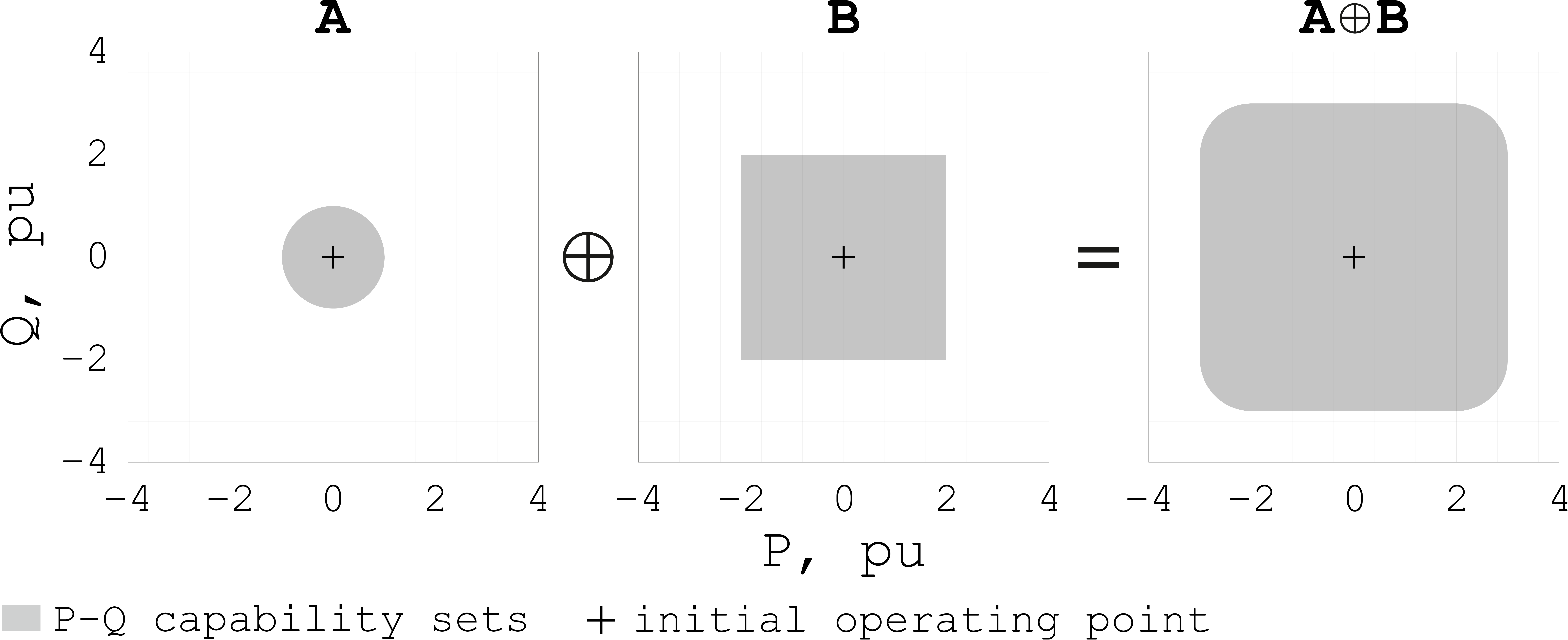}
    \caption{Example: Minkowski addition of P-Q capability sets of two flexible units. Network constraints are not considered.}
    \label{Fig: Minkowski_example1}
\end{figure}

An example of a Minkowski addition of two P-Q capability sets is shown in Fig.~\ref{Fig: Minkowski_example1}. The figure shows that the aggregated flexibility can be estimated as the boundary of $A \oplus B$. This example will be recalled later in this section to illustrate the contributions of units to various flexibility requests.

However, in the context of aggregated ADN flexibility, applying the Minkowski addition has several drawbacks, as the approach does not consider network constraints and the physics of power flows. Accordingly, currently established aggregated flexibility analysis approaches use OPF models to estimate the set of feasible operating points at a selected reference bus where the flexible power from different units can be aggregated, such as at the TSO/DSO interface \cite{Heleno2015,Silva2018,Contreras2018,Capitanescu2018,Lopez2021,Bolfek2021,Contreras2021}. Such models enable including network constraints (e.g., power balance constraints, power flow and voltage limits), DER constraints of flexibility provision, and power flow equations. Several different power flow formulations have been used in the literature to describe the flexibility of distribution networks, such as linearised OPF models \cite{Contreras2018}, nonlinear AC OPF models in polar and rectangular voltage coordinates \cite{Silva2018,Capitanescu2018}, the DistFlow OPF formulation for radial networks \cite{Lopez2021}. In this work, the latter formulation (the DistFlow model) is selected to estimate the flexibility of radial ADNs: it enables formulating network constraints as quadratic equations and is equivalent to the exact AC power flow equations for systems with radial topology, thus capturing the nonlinear nature of active and reactive power flows and voltage constraints \cite{Baran1989-1,Baran1989-2}. The model presented in \eqref{Model DistFlow: Pgen. output}-\eqref{Model DistFlow: x_activation_q} is modified by including power outputs of flexible units and binary variables corresponding to flexible power regulations. These variables will be used in logical constraints to forbid simultaneous consumption and production of flexible power for each unit, formulate coalitions of units, and study coordination issues of flexibility provision, such as the flexible power swap phenomenon.\footnote{Binary variables corresponding to flexible power regulations may be unnecessary for units with strictly increasing cost functions, as a single-period cost-minimising model for flexibility estimation will never provide a solution with a unit simultaneously consuming and producing power. However, it is advised to introduce these variables and logical constraints for models with multiple time periods or not cost-minimising objective functions.} The modified DistFlow model is a mixed integer quadratically constrained programming (MIQCP) problem, which can be solved with modern MIQCP solvers such as Gurobi 10.0.


\begin{model}[t]
\caption{{Modified DistFlow} \hfill [MIQCP]}
\label{Model1}
\begin{subequations} 
\vspace{-3\jot}
\begin{IEEEeqnarray}{lll}
    {\textbf{Variables:}} & \IEEEnonumber\\
    p_{n,g}, q_{n,g}  & n \in \mathcal{N},  g \in \mathcal{G} & \IEEEnonumber \\
    p_{nm}, q_{nm}  & (n,m) \in \mathcal{L} & \IEEEnonumber \\
    V_n \quad\, ({V_n}^2=w_n) & n \in \mathcal{N} & \IEEEnonumber \\
    i_{nm} \quad ({i_{nm}}^2=l_{nm})  & (n,m) \in \mathcal{L} & \IEEEnonumber \\    p^\uparrow_{n,f},p^\downarrow_{n,f},q^\uparrow_{n,f},q^\downarrow_{n,f}  & n \in \mathcal{N},  f \in \mathcal{F} & \IEEEnonumber \\    
    x_{n,f} \in \{0,1\}  & n \in \mathcal{N},  f \in \mathcal{F} & \IEEEnonumber\\    x^{p\uparrow}_{n,f},x^{p\downarrow}_{n,f},x^{q\uparrow}_{n,f},x^{q\downarrow}_{n,f} \in \{0,1\} & n \in \mathcal{N}, f \in \mathcal{F} & \IEEEnonumber \\    \mathfrak{x}^{p\uparrow},\mathfrak{x}^{p\downarrow},\mathfrak{x}^{q\uparrow},\mathfrak{x}^{q\downarrow} \in \{0,1\} & \IEEEnonumber
    \vspace{1\jot}\\
    {\textbf{Constraints:}} & \IEEEnonumber\\
        p^{\text{min}}_{n,g} \leq p_{n,g} \leq p^{\text{max}}_{n,g} &\forall n \in \mathcal{N}, g \in \mathcal{G} \qquad \label{Model DistFlow: Pgen. output} \\
        q^{\text{min}}_{n,g} \leq q_{n,g} \leq q^{\text{max}}_{n,g} &\forall n \in \mathcal{N}, g \in \mathcal{G} \label{Model DistFlow: Qgen. output} \\
        p_{nm} = \smashoperator{\sum_{d \in \mathcal{D}, g \in \mathcal{G}}} \big{(}p_{m,d} - p_{m,g} \big{)} \label{Model DistFlow: Pflow}\\
        \qquad + \smashoperator{\sum_{f \in \mathcal{F}}} \big{(}x^{p\downarrow}_{m,f} p^\downarrow_{m,f} - x^{p\uparrow}_{m,f} p^\uparrow_{m,f}\big{)} \IEEEnonumber\\ \qquad + \Re(Z_{nm})l_{nm} +  \smashoperator{\sum_{(m,k) \in \mathcal{L}}}p_{mk} \quad &\forall (n,m) \in \mathcal{L} \IEEEnonumber\\
        q_{nm} = \smashoperator{\sum_{d \in \mathcal{D}, g \in \mathcal{G}}} \big{(}q_{m,d} - q_{m,g} \big{)} \label{Model DistFlow: Qflow}\\
        \qquad + \smashoperator{\sum_{f \in \mathcal{F}}} \big{(}x^{q\downarrow}_{m,f} q^\downarrow_{m,f} - x^{q\uparrow}_{m,f} q^\uparrow_{m,f}\big{)} \IEEEnonumber\\ \qquad + \Im(Z_{nm})l_{nm} + \smashoperator{\sum_{(m,k) \in \mathcal{L}}}q_{mk}  &\forall (n,m) \in \mathcal{L} \IEEEnonumber\\
        w_m = w_n + \abs{Z_{nm}}^2 l_{nm} \label{Model DistFlow: voltage relation}\\ \qquad - 2\big{(}\Re(Z_{nm})p_{nm} + \Im(Z_{nm})q_{nm}\big{)} \enskip &\forall (n,m) \in \mathcal{L} \IEEEnonumber\\
        p_{nm}^2 + q_{nm}^2 = l_{nm}w_n  &\forall (n,m) \in \mathcal{L} \quad \label{Model DistFlow: apparent power} \\
        p_{nm}^2 + q_{nm}^2 \leq (\mathcal{S}_{nm}^\text{max})^2  &\forall (n,m) \in \mathcal{L} \quad \label{Model DistFlow: line limits} \\
        (V_n^\text{min})^2 \leq w_n \leq (V_n^\text{max})^2  &\forall n \in \mathcal{N} \quad \label{Model DistFlow: voltage limits}\\
        0 \leq p^\uparrow_{n,f} \leq \mathfrak{x}^{p\uparrow} x^{p\uparrow}_{n,f}{p^{\uparrow\text{max}}_{n,f}} &\forall n \in \mathcal{N}, f \in \mathcal{F} \label{Model DistFlow: p_up}\\
        0 \leq p^\downarrow_{n,f} \leq \mathfrak{x}^{p\downarrow} x^{p\downarrow}_{n,f}{p^{\downarrow\text{max}}_{n,f}} &\forall n \in \mathcal{N}, f \in \mathcal{F} \label{Model DistFlow: p_down}\\
        0 \leq q^\uparrow_{n,f} \leq \mathfrak{x}^{q\uparrow} x^{q\uparrow}_{n,f}{q^{\uparrow\text{max}}_{n,f}} &\forall n \in \mathcal{N}, f \in \mathcal{F} \label{Model DistFlow: q_up}\\
        0 \leq q^\downarrow_{n,f} \leq \mathfrak{x}^{q\downarrow} x^{q\downarrow}_{n,f}{q^{\downarrow\text{max}}_{n,f}} &\forall n \in \mathcal{N}, f \in \mathcal{F} \label{Model DistFlow: q_down}\\
        x^{p\uparrow}_{n,f} + x^{p\downarrow}_{n,f} \leq x_{n,f} &\forall n \in \mathcal{N}, f \in \mathcal{F} \label{Model DistFlow: x_activation_p}\\
        x^{q\uparrow}_{n,f} + x^{q\downarrow}_{n,f} \leq x_{n,f} &\forall n \in \mathcal{N}, f \in \mathcal{F} \label{Model DistFlow: x_activation_q}
        \vspace{-1\jot}
\end{IEEEeqnarray}
\end{subequations}
\end{model}

The variables of the model are the active and reactive power of conventional controllable generators, $p_{n,g}$, $q_{n,g}$, power flows, $p_{nm}$, $q_{nm}$, bus voltages, $V_n$, and branch currents, $i_{nm}$, where $n \in \mathcal{N}$ and $(n,m) \in \mathcal{L}$ are the sets of network buses and lines. The power of each flexible unit $f \in \mathcal{F}$ is given by its available P-Q upward and downward regulation capacities indicated by the corresponding arrows. That is, units can produce or consume flexible power, subject to their initial operating points. Binary variables $x_{n,f}$ define the availability status of units: units with $x_{n,f}=0$ cannot provide any flexible power regulation. Binary variables such as $x^{p\uparrow}_{n,f}$ correspond to P-Q upward and downward regulations of each unit, while variables such as $\mathfrak{x}^{p\uparrow}$ control regulation decisions for all available units together. Active and reactive power of generators is limited in \eqref{Model DistFlow: Pgen. output}-\eqref{Model DistFlow: Qgen. output}. Equations \eqref{Model DistFlow: Pflow}-\eqref{Model DistFlow: Qflow} define power flows between buses $n$ and $m$ in a radial grid, where $\Re(Z_{nm})$ and $\Im(Z_{nm})$ are the real and imaginary parts of the branch impedance, and $p_{m,d}$ and $q_{m,d}$ are nodal loads. The voltage relation between buses $n$ and $m$ is defined by \eqref{Model DistFlow: voltage relation}. Equation \eqref{Model DistFlow: apparent power} determines the relation between branch power flows and currents. The apparent power flow limits and nodal voltage limits are imposed in \eqref{Model DistFlow: line limits}, \eqref{Model DistFlow: voltage limits}. The P-Q capability of each flexible unit is defined in \eqref{Model DistFlow: p_up}-\eqref{Model DistFlow: q_down}, where upward and downward regulation capacities are interrelated with the corresponding binary decision variables. Finally, logical constraints \eqref{Model DistFlow: x_activation_p}-\eqref{Model DistFlow: x_activation_q} state that, if available, units can perform only one of the regulations (either produce or consume flexible power).


In this work, the modified DistFlow model serves as the basis for flexibility tracing, ranking, and valuation in radial distribution networks. This formulation also enables analysing the nonlinear flexible unit dispatch (and potential unit coordination issues) induced by the OPF quadratic terms, e.g., in \eqref{Model DistFlow: voltage relation}-\eqref{Model DistFlow: voltage limits}.
The models introduced below in this section will be iteratively using the DistFlow variables and constraints defined in \eqref{Model DistFlow: Pgen. output}-\eqref{Model DistFlow: x_activation_q} to approximate the flexibility area boundary, minimise the cost of flexibility services, and formulate cooperative games among flexible units. Before introducing these models, it is important to define the aggregated power at the selected reference bus, e.g., a primary substation or TSO/DSO interface. For bus $n^{\text{ref}}$, the aggregated active and reactive power is the sum of power injections (power flows through connected branches and power regulations of flexible units at this bus):
\begin{subequations} 
\begin{IEEEeqnarray}{lll}
    P_{n} = \smashoperator{\sum_{(n,m) \in \mathcal{L}}} p_{nm} + \smashoperator{\sum_{f \in \mathcal{F}}} \big{(}x^{p\downarrow}_{n,f} p^\downarrow_{n,f} - x^{p\uparrow}_{n,f} p^\uparrow_{n,f}\big{)} \quad & n = n^{\text{ref}} \qquad \label{Aggregated power: P}\\
    Q_{n} = \smashoperator{\sum_{(n,m) \in \mathcal{L}}} q_{nm} + \smashoperator{\sum_{f \in \mathcal{F}}} \big{(}x^{q\downarrow}_{n,f} q^\downarrow_{n,f} - x^{q\uparrow}_{n,f} q^\uparrow_{n,f}\big{)} \quad & n = n^{\text{ref}} \qquad \label{Aggregated power: Q}
\end{IEEEeqnarray}
\end{subequations}


Then, the boundary estimation of the aggregated flexibility area can be formulated as model \eqref{Model boundary: objective}-\eqref{Model boundary: epsilon q}. Objective function \eqref{Model boundary: objective} minimises or maximises the network’s power consumption at the reference bus, where coefficients $\alpha^p_n$ and $\alpha^q_n$ control the optimisation directions in the P-Q space. Constraints of the modified DistFlow model and the aggregated power equations are introduced in \eqref{Model boundary: Distflow}, \eqref{Model boundary: Pn_ref Qn_ref}. In \eqref{Model boundary: epsilon p}, \eqref{Model boundary: epsilon q}, the aggregated active and reactive power is limited by $\epsilon$-intervals that are used to produce the piece-wise linear approximation of the flexibility area boundary with operating points ($P_{n}^\prime$,$Q_{n}^\prime$).
\begin{model}[t]
\caption{{Flexibility boundary estimation} \hfill [MIQCP]}
\label{Model2}
\begin{subequations} 
\label{Mod: TEP1}
\vspace{-3\jot}
\begin{IEEEeqnarray}{lll}
    {\textbf{Objective:}} & \IEEEnonumber\\
    \min \enskip \alpha^p_n P_{n} + \alpha^q_n Q_{n} \quad & n = n^{\text{ref}} \quad \label{Model boundary: objective} \vspace{1\jot}\\
    {\textbf{Constraints:}} & \IEEEnonumber\\
    \text{modified DistFlow model \eqref{Model DistFlow: Pgen. output}-\eqref{Model DistFlow: x_activation_q}} \label{Model boundary: Distflow}\\
    \text{aggregated flexibility $P_{n}$, $Q_{n}$ \eqref{Aggregated power: P}-\eqref{Aggregated power: Q}} \qquad & n = n^{\text{ref}} \quad \label{Model boundary: Pn_ref Qn_ref}\\
    P_{n}^\prime - \epsilon^p_n \leq P_{n} \leq  P_{n}^\prime + \epsilon^p_n & n = n^{\text{ref}} \label{Model boundary: epsilon p}\\
    Q_{n}^\prime - \epsilon^q_n \leq Q_{n} \leq  Q_{n}^\prime + \epsilon^q_n & n = n^{\text{ref}}\label{Model boundary: epsilon q}
    \vspace{-1\jot}
\end{IEEEeqnarray}
\end{subequations}
\end{model}

The optimisation model \eqref{Model boundary: objective}-\eqref{Model boundary: epsilon q} can be solved iteratively to approximate the boundary of the network flexibility area at the reference bus with the desired level of granularity.\footnote{The discussion of the flexibility areas construction algorithms can be found in \cite{Lopez2021,Bolfek2021,Contreras2021}.} 
However, this approach does not provide additional information on the operating conditions within the area, e.g., what units have to be activated, what level of coordination is required to reach certain operating points, and how to remunerate units for providing different flexibility requests. Moreover, the boundary-estimation model cannot incorporate additional metrics of flexibility, such as the cost and economic surplus of flexible power provision. These issues are addressed with more advanced models and tools in the following subsections.

\subsection{Network Flexibility as a Cost-minimising OPF Problem}
For each feasible operating point (a flexibility service request), the aggregated flexibility of a distribution network can be characterised by the costs associated with the flexible power of available units \cite{Silva2018,Riaz2021}. A cost-minimisation model can identify the cheapest units to be activated for providing a specific flexible power request, naturally capturing their contributions to the network aggregated flexibility. Given a flexibility request ($P_{n}^\prime$,$Q_{n}^\prime$), i.e., a new operating point requested for a flexibility service at the reference bus, the cost-minimising OPF problem can be formulated as presented in \eqref{Model cost: objective}-\eqref{Model cost: Q request}. Objective function \eqref{Model cost: objective} minimises the costs of upward and downward regulations for all flexible units available in the network. Note that the units' cost functions $C_{n,f}$ can set different costs for producing and consuming flexible active or reactive power. Constraints of the modified DistFlow model and the aggregated power equations are introduced in \eqref{Model cost: DistFlow}, \eqref{Model cost: Pn_ref Qn_ref}. Aggregated active and reactive power at the new operating point requested is defined in \eqref{Model cost: P request}, \eqref{Model cost: Q request}.

\begin{model}[t]
\caption{{Cost minimisation of flexibility requests} \hfill [MIQCP]}
\label{Model3}
\begin{subequations} 
\vspace{-3\jot}
\begin{IEEEeqnarray}{lll}
    {\textbf{Objective:}} & \IEEEnonumber\\
    \min \enskip  \smashoperator{\sum_{n \in \mathcal{N}}} \smashoperator{\sum_{f \in \mathcal{F}}} C_{n,f}(p^\uparrow_{n,f},p^\downarrow_{n,f},q^\uparrow_{n,f},q^\downarrow_{n,f}) \quad \label{Model cost: objective} \vspace{1\jot}\\
    {\textbf{Constraints:}} & \IEEEnonumber\\
    \text{modified DistFlow model \eqref{Model DistFlow: Pgen. output}-\eqref{Model DistFlow: x_activation_q}} \label{Model cost: DistFlow}\\
    \text{aggregated flexibility $P_{n}$, $Q_{n}$ \eqref{Aggregated power: P}-\eqref{Aggregated power: Q}} \qquad & n = n^{\text{ref}} \quad \label{Model cost: Pn_ref Qn_ref}\\
    P_{n} = P_{n}^\prime & n = n^{\text{ref}} \label{Model cost: P request}\\
    Q_{n} = Q_{n}^\prime & n = n^{\text{ref}} \label{Model cost: Q request}
    \vspace{-1\jot}
\end{IEEEeqnarray}
\end{subequations}
\end{model}

The cost-minimising model \eqref{Model cost: objective}-\eqref{Model cost: Q request} enables to analyse the aggregated network flexibility at any feasible operating point, both at the flexibility area boundary and within the boundary.
A solution to \eqref{Model cost: objective}-\eqref{Model cost: Q request} provides information on the components of flexible power for a given flexibility request, i.e., the optimal flexible power outputs of the units, $p^{\uparrow*}_{n,f}$, $p^{\downarrow*}_{n,f}$, $q^{\uparrow*}_{n,f}$, and $q^{\downarrow*}_{n,f}$. Therefore, this model can serve as a tool for tracing, ranking, and valuation of network flexibility, where the cheapest flexible units get activated subject to network constraints.
Moreover, the cost-based formulation enables estimating additional metrics to characterise network flexibility. For example, the optimised objective function \eqref{Model cost: objective} provides the total minimum cost of meeting a flexibility request:
\begin{IEEEeqnarray}{lll}
    \label{Flex_total_cost}
    {F^\text{cost}} = \smashoperator{\sum_{n \in \mathcal{N}}} \smashoperator{\sum_{f \in \mathcal{F}}} C_{n,f}(p^{\uparrow*}_{n,f},p^{\downarrow*}_{n,f},q^{\uparrow*}_{n,f},q^{\downarrow*}_{n,f})
\end{IEEEeqnarray}


Assume that there exists a TSO-DSO flexibility market in which TSO pays DSO for active and reactive power control at the reference bus $n^\text{ref}$, $\Delta P_n = P_n^\prime - P_n^0$, $\Delta Q_n = Q_n^\prime - Q_n^0$.\footnote{This market assumption, used to estimate the economic surplus of flexibility provision, may not be true for some cases and countries. Nevertheless, the presented ideas of flexibility ranking and valuation are general and can be adapted to different TSO-DSO coordination schemes.} If these payments can be described by a market price function $\pi$, then the economic surplus of flexibility provision is the difference between the payment and the optimised cost of flexible power:
\begin{IEEEeqnarray}{lll}
    \label{Flex_surplus}
    {F^\text{surplus}} = \pi_n (\Delta P_n, \Delta Q_n)- {F^\text{cost}} \qquad & n = n^{\text{ref}} \quad
\end{IEEEeqnarray}


These metrics are useful for interpreting network flexibility since they capture both the flexible power outputs of the units and their economic impacts, such as the cost and the economic surplus of power provision. However, the cost-minimising OPF problem formulation is biased as it selects only the cheapest flexible units and does not consider all possible contributions of other units to network flexibility. In this regard, the following subsection introduces the cooperative game formulation of network flexibility, where different possible combinations of flexible units and their contributions are included in the tracing, ranking, and valuation mechanisms.

\subsection{Network Flexibility as a Cooperative Game among Flexible Units}\label{Subsection: flexibility as a cooperative game}
The flexibility of a distribution network with multiple flexible units has an inherent combinatorial nature: some flexible power requests can be provided by any of the available flexible units, whereas other requests require different combinations of units (or even all the units) to be activated.
Units with higher potential contributions to a flexibility request are more critical for the operation of the DSO, and it can be argued that these units should receive additional availability payments for ensuring the security of flexibility services.
Therefore, a comprehensive analysis of network flexibility should account for possible combinations of units and their contributions to the flexible power provision. Such an analysis can be performed using the well-established tools from Cooperative Game Theory. This section introduces the main game-theoretic concepts that can be used for tracing, ranking, and valuating flexibility in distribution networks. A more thorough description of cooperative games and Cooperative Game Theory solution concepts can be found in \cite{Maschler2013,Chalkiadakis2011}.

Assuming that simultaneous activation of flexible units located in a distribution network brings some value of cooperation, which can be divided and transferred between the units, a cooperative game $(N;v)$ with transferable utility can be defined as follows:\footnote{
Note that the representation of flexible units as players joining coalitions is an abstract notion used to describe the combinatorial problem of joint flexible power provision by multiple units. That is, units do not take decisions in the formation of coalitions or compete for joining them. In this work, cooperative games are used to allocate the values between flexible units and rank them. It is not implied that units have to join coalitions during this allocation and ranking process.}

\begin{itemize}
	\item $N$ is a finite set of players (flexible units available in the network for flexible power provision). A subset of $N$ is called a coalition. The largest possible coalition containing all players is called the grand coalition. As further demonstrated in this work, the grand coalition provides the greatest amount of available flexible power. The collection of all coalitions is denoted by $2^N$.
	\item $v$ $:$ ${2^N} \rightarrow \mathbb{R}$ is the characteristic function associating each coalition $S$ with a real number $v(S)$, a metric describing the value of a coalition. In this work, various metrics will be used to describe the value of coalitions, such as the limits of the aggregated network flexibility, the cost of flexible power provision, and its economic surplus.
\end{itemize}

In the context of the P-Q aggregated flexibility provision at the reference bus, the formation of a coalition $S$ and its value can be defined as presented in \eqref{Model game: objective}-\eqref{Model game: binaries-coalitions}. Objective function \eqref{Model game: objective} maximises flexible power regulation provided by a coalition. This regulation can correspond to flexible production or consumption depending on the power control requested, i.e., the signs of P-Q deviations, $\text{sgn}(\Delta P_{n})$ and $\text{sgn}(\Delta Q_{n})$. Constraints of the modified DistFlow model and the aggregated power equations are introduced in \eqref{Model game: DistFlow}, \eqref{Model game: Pn_ref Qn_ref}.
In \eqref{Model game: delta P}, \eqref{Model game: delta Q}, changes in the aggregated flexible power with respect to the initial operating point ($P_{n}^0$,$Q_{n}^0$) are limited by the requested flexible power, i.e., a coalition of units will not provide more flexible power than requested from the DSO. The direction of the flexible power regulation is given in \eqref{Model game: cos_phi} by the power factor (P-Q ratio) of the flexibility request. Finally, \eqref{Model game: binaries-coalitions} defines the formation of coalitions among flexible units.
\begin{model}[t]
\caption{{Coalitional analysis of flexibility requests} \hfill [MIQCP]}
\label{Model4}
\begin{subequations} 
\vspace{-3\jot}
\begin{IEEEeqnarray}{lll}
    {\textbf{Objective:}} & \IEEEnonumber\\
    \max \enskip \text{sgn}(\Delta P_{n})P_{n} +  \text{sgn}(\Delta Q_{n})Q_{n} & n = n^{\text{ref}} \qquad\enskip \label{Model game: objective} \vspace{1\jot}\\
    {\textbf{Constraints:}} & \IEEEnonumber\\
    \text{modified DistFlow model \eqref{Model DistFlow: Pgen. output}-\eqref{Model DistFlow: x_activation_q}} \label{Model game: DistFlow}\\
    \text{aggregated flexibility $P_{n}$, $Q_{n}$ \eqref{Aggregated power: P}-\eqref{Aggregated power: Q}} \qquad & n = n^{\text{ref}} \quad \label{Model game: Pn_ref Qn_ref}\\
    (P_{n} - P_{n}^0)\text{sgn}(\Delta P_{n}) \leq \Delta P_{n} \text{sgn}(\Delta P_{n}) & n = n^{\text{ref}} \label{Model game: delta P}\\
    (Q_{n} - Q_{n}^0)\text{sgn}(\Delta Q_{n}) \leq \Delta Q_{n} \text{sgn}(\Delta Q_{n}) & n = n^{\text{ref}} \label{Model game: delta Q}\\
    (P_{n} - P_{n}^0) = (Q_{n} - Q_{n}^0) \cos(\phi)^\prime & n = n^{\text{ref}} \label{Model game: cos_phi}\\
    x_{n,f} = \begin{dcases*}
        1  & \text{if unit is part of coalition $S$}\\
        0  &\text{otherwise}. 
        \end{dcases*} & \begin{aligned}\forall n \in \mathcal{N}, \\f \in \mathcal{F} \end{aligned}\label{Model game: binaries-coalitions}
    \vspace{-1\jot}
\end{IEEEeqnarray}
\end{subequations}
\end{model}

Any coalition $S$ can be characterised by the aggregated flexibility regulations obtained from model \eqref{Model game: objective}-\eqref{Model game: binaries-coalitions} or by the cost and economic surplus of these regulations, ${F^\text{cost}(S)}$ and ${F^\text{surplus}(S)}$, as defined in \eqref{Flex_total_cost}, \eqref{Flex_surplus}. Then, the cooperative game formulation can be used to allocate the value of the grand coalition, $v(N)$, among the units, thus ranking them according to the selected metric. The crucial measure of a player's impact in a cooperative game is the marginal contributions to the coalitions the player can join. The marginal contribution to coalition $S$ by player $i$ is estimated as the difference in the value of the coalition with and without the player:
\begin{IEEEeqnarray}{lll}
    \label{CGT: MC_i}
    {MC(S)}_i = v(S \cup \{i\})-v(S) \quad \forall i \in S \quad \forall S \subseteq N 
\end{IEEEeqnarray} 

Then, the allocation of the grand coalition value, $v(N)$, to player $i$ can be found as the weighted average of player's marginal contributions to all possible coalitions, as defined by the Shapley value formula:
\begin{IEEEeqnarray}{l}
    \label{CGT: Shapley}
    Sh_i=\sum_{S \subseteq N \setminus \{i\} } \frac{|S|!(|N|-|S|-1)!}{|N|!} {MC(S)}_i
    \enskip\enskip\enskip
\end{IEEEeqnarray}

The weight of each coalition depends on the number of players in the coalition, $|S|$, and the total number of players in a cooperative game, $|N|$. The Shapley value has been acknowledged as a useful tool in cost allocation mechanisms, data analysis, and power systems research \cite{Churkin2021review,Banez-Chicharro2017,Hasan2018,Kristiansen2018,Han2019,Hupez2021,Fleischhacker2022,Anibal2022,Cremers2023,Ai2022,Goncalves2021}. Several desirable properties of the Shapley value make it suitable for tracing, ranking, and valuation of flexibility in distribution networks. The \textit{symmetry} property guarantees that two identical flexible units that bring equal contributions to coalitions will always be allocated the same value. The \textit{null player} property states that a unit that contributes nothing to any coalition will not receive a share of the grand coalition value. The \textit{efficiency} property requires that the sum of the values allocated to all players is equal to the value of the grand coalition. In this work, the Shapley value is exploited to analyse the contributions of flexible units to different flexibility requests, thus tracing the flexible power and ranking units by their criticality. It is also demonstrated that the Shapley value can be used as a valuation mechanism for flexible units.

\begin{figure}
    \centering
    \includegraphics[width=\columnwidth]{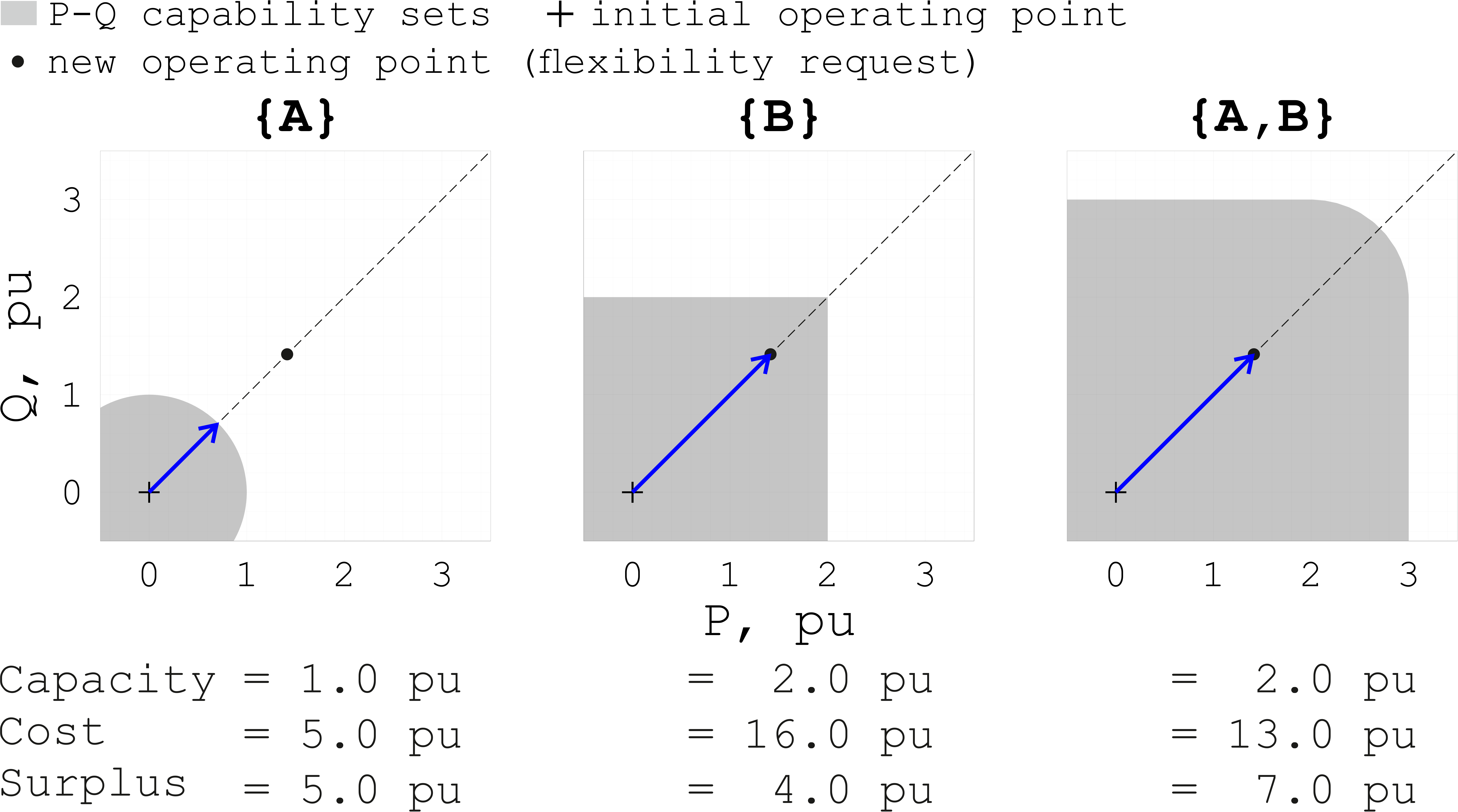}
    \caption{Estimating the contributions of flexible units in the 2-player Minkowski addition example. Network constraints are not considered.}
    \label{Fig: Minkowski_example2}
\end{figure}

Application of the coalitional model \eqref{Model game: objective}-\eqref{Model game: binaries-coalitions} and the Shapley value to aggregated flexibility requests is illustrated based on the Minkowski addition example with two flexible units, as shown in Fig.~\ref{Fig: Minkowski_example2}. In this example, two flexible units (players) A and B can form only three possible coalitions: \{A\}, \{B\}, \{A,B\}. Each coalition has different flexibility limits defined by the Minkowski addition \eqref{Eq: Minkowski}. The dashed line indicates the optimisation direction for coalitions towards the new operating point (flexibility request). Coalition \{B\} and \{A,B\} can fully provide the requested flexibility, while coalition \{A\} can provide it only partially. It is assumed that unit A provides flexibility at a cost of 5 p.u., unit B has a cost of 8 p.u., and the payment to DSO for this flexibility request (price $\pi$) is 10 p.u. Considering these assumptions and the flexibility limits of the coalitions, the cost and economic surplus for each coalition can be calculated as shown in Fig.~\ref{Fig: Minkowski_example2}. To analyse the ranking and valuation of the units in this example, Tables \ref{table: solutions least-cost} and \ref{table: solutions Shapley} show the allocation solutions provided by the cost-minimising model and the Shapley value.

\begin{table}[b]
\caption{Least-cost Allocation Solutions for the 2-player \\Minkowski Addition Example}
\centering
\begin{tabular}{@{}lccc@{}}
\toprule
 \multirow{2}{*}{Player} & \multicolumn{3}{c}{Allocation solutions by the cost-minimising model, in p.u.}\\
  \cmidrule(l){2-4} 
 & Capacity & Cost & Economic surplus\\
\midrule
\vspace{1\jot}
 A & 1.0 (50.0\%) & 5.0 (38.5\%) & 5.0 (71.4\%)\\
 B & 1.0 (50.0\%) & 8.0 (61.5\%) & 2.0 (28.6\%)\\
\bottomrule
\end{tabular}
\label{table: solutions least-cost}
\end{table}

\begin{table}[b]
\caption{Shapley Value Allocation Solutions for the 2-player Minkowski Addition Example}
\centering
\begin{tabular}{@{}lccc@{}}
\toprule
 \multirow{2}{*}{Player} & \multicolumn{3}{c}{Allocation solutions by the Shapley value, in p.u.}\\
  \cmidrule(l){2-4} 
 & Capacity & Cost & Economic surplus\\
\midrule
\vspace{1\jot}
 A & 0.5 (25.0\%) & 1.0 (7.7\%) & 4.0 (57.1\%)\\
 B & 1.5 (75.0\%) & 12.0 (92.3\%) & 3.0 (42.9\%)\\
\bottomrule
\end{tabular}
\label{table: solutions Shapley}
\end{table}

The cost-minimising model identifies units A and B as equally critical since in \{A,B\} each of them provides 1.0 p.u. of flexible power regulation in the required direction. Unit B, however, has a significantly higher P-Q capability and is clearly more critical for flexibility service provision. The Shapley value captures potential contributions by unit B to the flexible power regulation and identifies it as 75\% critical for the requested flexibility service. It follows that the Shapley value can be a preferable tool for tracing the aggregated network flexibility, ranking the criticality of units, and estimating the diversification of flexible resources. As a remuneration mechanism, the cost-minimising model allocates 71\% of the economic surplus to unit A since it provides the same flexible power regulation at a lower cost. This remuneration does not consider contributions of unit B to the security of flexibility provision: unit B gets paid only for the 1.0 p.u. of flexible power delivered, and not for its availability and potential contributions. In contrast, the Shapley value includes the contributions of unit B in different coalitions and allocates the economic surplus more evenly. Thus, the Shapley value can be used as a remuneration mechanism that inherently includes availability and delivery (utilisation) payments. In Section~\ref{Section: results}, the application of the coalitional model \eqref{Model game: objective}-\eqref{Model game: binaries-coalitions} and the Shapley value in flexibility ranking and valuation will be demonstrated based on a realistic distribution network.

\subsection{Effects of Unit Coordination and Nonlinearity of Flexible Power Provision}\label{Subsection: coordination and nonlinearity}
The models introduced above identify the optimal flexible unit dispatch that maximises the aggregated network flexibility or minimises the costs of flexible power regulations. However, such solutions rely on the assumption of perfect unit coordination, i.e., units can instantly exchange information and take any actions to jointly regulate the network operation and flexibility provision. For example, due to the nonlinear constraints of the power flow model \eqref{Model DistFlow: Pgen. output}-\eqref{Model DistFlow: x_activation_q}, some units have to perform active network management, fast flexible power regulations, and voltage control. In reality, unit coordination might be not perfect, and intertemporal constraints may limit changes in their flexible power regulation.

There are several potential issues related to unit coordination and the nonlinearity of flexible power provision that may be undesirable for DSOs. First, in some operating points, units can exchange (swap) flexible power to alleviate network constraints and maximise network flexibility. 
{\color{black}DSOs can provide aggregated flexibility in such regimes only under the assumption of perfect unit coordination. That is, if flexible units and the DSO exchange information and centrally coordinate actions, e.g., to determine which units should be managing network constraints and which units provide cheap flexibility services. However, the assumption of perfect flexible unit coordination may be unrealistic (unless the units are owned or operated by the same actor). Some units may only be partially controllable by the DSO and would generally not exchange all information with other units.}
In this regard, the following unit coordination constraints can be included in the models to disable flexible power swaps:
\begin{subequations} 
\begin{IEEEeqnarray}{lll}
    \mathfrak{x}^{p\uparrow} + \mathfrak{x}^{p\downarrow} <= 1 \label{Power swap constraint: p}\\
    \mathfrak{x}^{q\uparrow} + \mathfrak{x}^{q\downarrow} <= 1 \label{Power swap constraint: q}
\end{IEEEeqnarray}
\end{subequations} 

These no-swap constraints limit flexible power regulation decisions globally for all units. That is, each unit can only produce or consume flexible power. There is no pair of units that consume and produce flexible power, thus exchanging it.

Second, units may need to shift their flexible power regulations rapidly between close operating points. 
{\color{black}Such rapid nonlinear shifts in the optimal flexible unit dispatch are justified by both economic reasons (differences in the flexible units’ cost functions) and network constraints (impact of thermal and voltage constraints on the flexible unit operations). However, in reality, flexible units might not have sufficient ramp capabilities to follow such a complex dispatch with rapid nonlinear changes. 
The presence of slow units can reduce the economic efficiency of the flexibility services or even put the system operation at risk. Therefore, as DSOs increasingly use flexibility, it becomes necessary to characterise the nonlinearity of the optimal flexible unit dispatch and identify risky areas that may require fast flexible power regulations.}

{\color{black}Measuring the nonlinearity of a mathematical optimisation model is a complex task with many nuances. The general idea of nonlinearity assessment is to identify the degree of deviation from linearity in the objective function, constraints, or variables. 
In a simple problem, rapid nonlinear changes in variables can be identified by analysing the derivatives of the corresponding functions. However, in the complex context of the optimal flexible unit dispatch, the functions of units' flexible power are not known a piori. The optimal flexible power regulations for each unit can be only found for a given set of operating points by solving the cost-minimising OPF model \eqref{Model cost: objective}-\eqref{Model cost: Q request}. Therefore, the nonlinear behaviour of flexible units can be characterised by a nonlinearity metric derived by comparing solutions for a given set of operating points.}
Considering a feasible P-Q space discretised with $k$ operating points, the nonlinearity in flexible power provision can be measured as the maximum changes in flexible power regulations of each unit between neighbouring points:
\begin{IEEEeqnarray}{lll}
    \label{Nonlinearity max}
    \overline{L} = \max_{n \in \mathcal{N}, f \in \mathcal{F}}\Big{\{} \max_{k \in K}\big{\{} \\
    \qquad \big{(}\max(p^\uparrow_{n,f,k}-p^\downarrow_{n,f,k})  - \min(p^\uparrow_{n,f,k}-p^\downarrow_{n,f,k}) \big{)}, \IEEEnonumber \qquad\\
    \qquad \big{(}\max(q^\uparrow_{n,f,k}-q^\downarrow_{n,f,k}) - \min(q^\uparrow_{n,f,k}-q^\downarrow_{n,f,k}) \big{)} \enskip
     \big{\}}  \Big{\}} \IEEEnonumber \quad
\end{IEEEeqnarray} 

Then, the nonlinearity factor can be introduced as the ratio between the maximum flexible power shift and the distance between the points (the discretisation grid step):
\begin{IEEEeqnarray}{lll}
    \label{Nonlinearity factor}
    \Delta{L} = \frac{\overline{L}}{\abs{\Delta k}}
\end{IEEEeqnarray} 

\begin{figure}
    \centering
    \includegraphics[width=0.45\columnwidth]{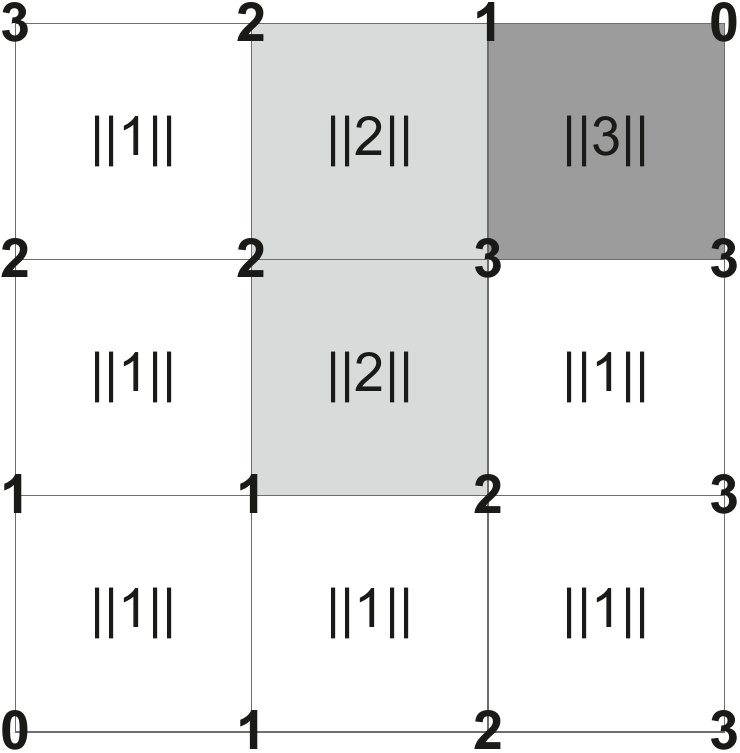}
    \caption{Example of the nonlinearity assessment algorithm for a 16-point grid. Values at the vertices denote unit's flexible power at the operating points. The absolute-value norms show maximum changes in the flexible power values for each segment and identify rapid nonlinear shifts in the flexible unit operation.}
    \label{Fig: nonlinearity_segments_explanation}
\end{figure}

{\color{black}Note that in this setting, the distance between considered operating points, $\abs{\Delta k}$, is the key parameter that determines the OPF solutions to be compared. The accuracy of a nonlinearity assessment algorithm depends on this distance and the number of operating points $k$. Therefore, DSOs can customise the algorithm's accuracy for various tasks, e.g., for analysing changes between a given set of P-Q operating points (which can be far from each other) or detecting potential rapid nonlinear changes in the flexible unit operation by considering a set of close operating points, among other applications.}\footnote{\color{black}Note that the proposed approach, similar to sensitivity analyses, identifies rapid nonlinear changes, but it does not explain them (e.g., does not indicate what constraints or parameters caused the nonlinear behaviour of units). In this regard, future research will explore approaches to capture the drivers for the nonlinearities (e.g., by incorporating dual variables of network constraints) to better inform DSOs on the aggregated DER flexibility provision.}
An example of nonlinearity assessment is shown in Fig.~\ref{Fig: nonlinearity_segments_explanation}, where a grid with 16 points illustrates the optimal operation of a flexible unit. The values between the points (in the segments of the area) denote the absolute value of the maximum changes between neighbouring points. In the white segments, the maximum changes are equal to the distance between the points. Thus, there are no rapid nonlinear shifts in the flexibility provision. However, in the grey segments, the flexible power of the unit changes rapidly, with the nonlinearity factor $\Delta{L}$ of 2 and 3. The DSO may want to avoid working in such areas as it can be difficult to ensure the economic efficiency and security of flexibility provision.

In the next section, the nonlinearity metric and the no-swap constraints will be used to analyse unit dispatch and coordination issues in a realistic distribution network.

\section{Simulation Results}\label{Section: results}

\subsection{Case Study: 33-bus Radial Distribution Network}
\begin{figure}
    \centering
    \includegraphics[width=0.9\columnwidth]{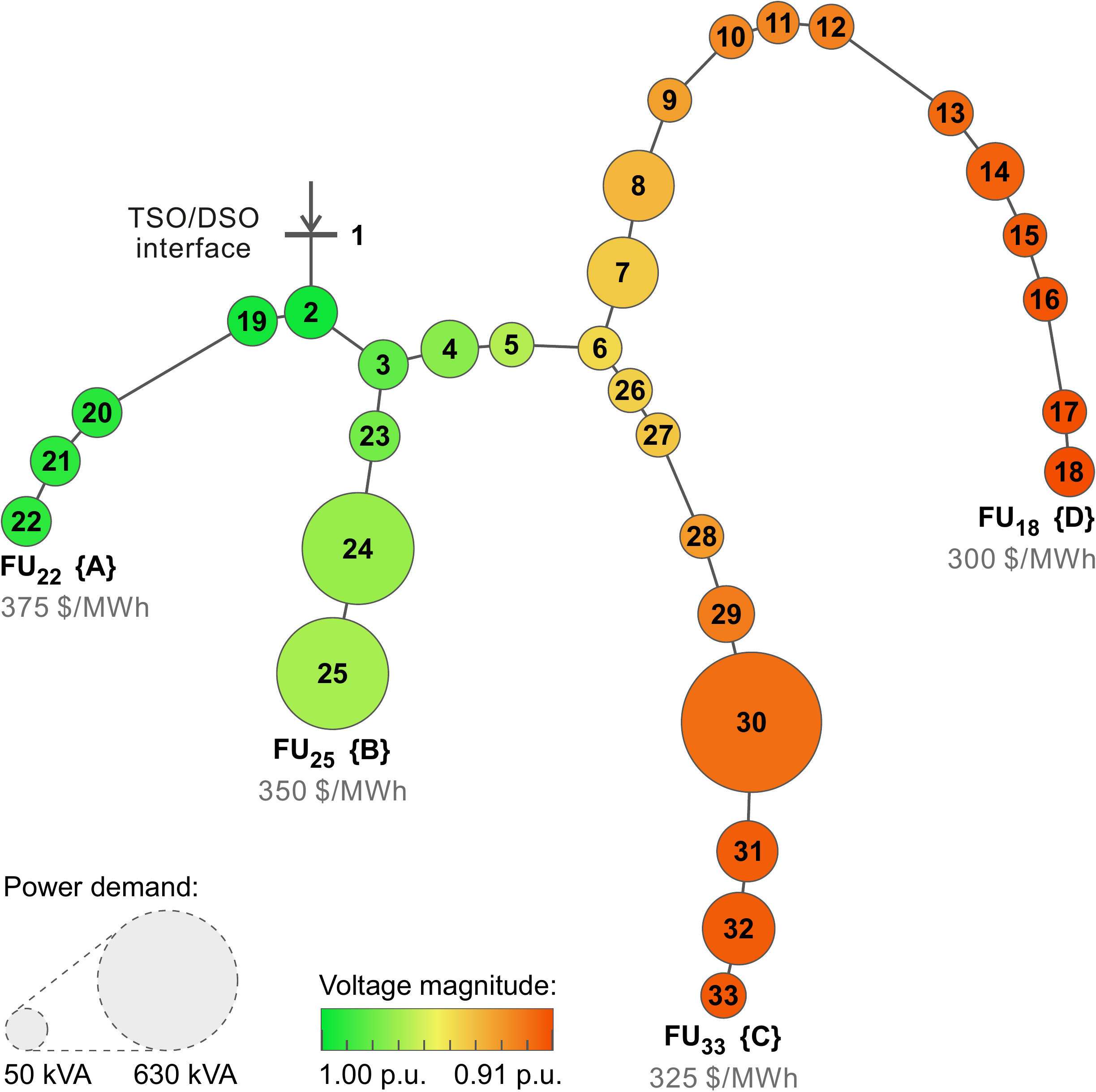}
    \caption{Case study: 33-bus radial distribution network with 4 flexible units.}
    \label{Fig: case33bw - scheme}
\end{figure}

The proposed methodology is demonstrated with the IEEE 33-bus test system, which is a 12.66kV radial distribution network \cite{Baran1989-2}. 
The total power demand of the network consumers is 3.7 MW and 2.3 MVAr.
The 33-bus system is visualised in Fig.~\ref{Fig: case33bw - scheme} with the force-directed graph layout algorithm \textit{ForceAtlas2} \cite{Jacomy2014}. The figure illustrates the network topology, power demands (as circles of different sizes), network voltage profile (nodes colouring), and electrical distances between the buses (lengths of the arcs). Note that the voltage levels at buses 18 and 33 are close to the lower limit of 0.9 p.u., which creates additional constraints on the network power consumption increase.\footnote{In the original 33-bus distribution system, the voltage limits are set to 0.9 p.u. and 1.1 p.u. However, more realistic cases can have tighter voltage constraints. For example, in the UK distribution networks, voltage deviations are limited to ±6\% of the nominal voltage. Regardless, the selection of the voltage limits does not alter the findings of this paper.} As further demonstrated by the simulations, such constraints can vastly affect the flexible power provided by units located in different parts of the network. 

To explore the constraints that voltage limits can have on the provision of flexibility, four flexible units are placed in the network (one at the end of each feeder). The four units are assumed to have identical P-Q capabilities: $p_{n,f}\in[-500,500]$ kW, $q_{n,f} \in [-500,500]$ kVAr, where $n \in \{18,22,25,33\}$ are the buses that the units are connected to. To ease the notations, the flexible units will be referred to as A, B, C, and D, as indicated in Fig.~\ref{Fig: case33bw - scheme}. The initial states of the units correspond to their not activated condition, i.e., units neither produce nor consume flexible power. Bus 1, the TSO-DSO interface, is considered as the reference bus, where the flexibility from the units is aggregated. 
To investigate the applicability of remuneration mechanisms, it is assumed that the units have different costs of providing flexible active power: 375, 350, 325, and 300 \$/MWh for units A, B, C, and D, respectively. Such costs are comparable to balancing market prices and are aligned with the cost assumptions in recent studies \cite{EID20191,Riaz2021}. The cost of flexible reactive power is assumed to be 50\% of the flexible active power cost for each unit. The price that TSO pays for flexible power in the TSO-DSO flexibility market is set at 400 \$/MWh and 200 \$/MVArh. Thus, the cost and the economic surplus of providing flexibility can be estimated as defined in \eqref{Flex_total_cost} and \eqref{Flex_surplus}.

All simulations presented in this section have been performed using JuMP 1.4.0 for Julia 1.6.1 programming language and Gurobi 10.0 solver.

\subsection{Structure of the Aggregated Network Flexibility}
First, the feasibility boundary estimation model \eqref{Model boundary: objective}-\eqref{Model boundary: epsilon q} is applied to analyse the aggregated network flexibility, which is the set of network feasible operating points when all flexible units are activated and fully coordinated.
This flexibility is displayed in Fig.~\ref{Fig: coalitional_structure} as the area reached by the set of units \{A,B,C,D\}, the grand coalition. Note that network flexibility at the TSO/DSO interface is more complex than a linear combination of units' P-Q capabilities (Minkowski addition). The resulting flexibility area has a nonlinear boundary due to the nonlinearities of the power flow model, such as the presence of power losses and voltage constraints.
The aggregated network flexibility can be decomposed into the P-Q capabilities of individual flexible units and their combinations. This decomposition corresponds to the coalitional structure of the cooperative game among units. A cooperative game with 4 players (units) consists of 15 possible coalitions, as illustrated in Fig.~\ref{Fig: coalitional_structure}. Model \eqref{Model boundary: objective}-\eqref{Model boundary: epsilon q} was iteratively  solved for each coalition to approximate the flexibility areas' boundaries. Synergy can be observed in the flexible unit coordination: units provide much more flexibility in large coalitions than in smaller coalitions or when being activated individually.


\begin{figure*}
    \centering
    \includegraphics[width=\textwidth]{coalitional_structure.pdf}
    \caption{Coalitional structure of the cooperative game among four flexible units. Each coalition is characterised by the aggregated network flexibility area in the P-Q space at the TSO/DSO interface. The markers correspond to the initial operating point of the network, while the coordinates represent the network's power consumption.
    }
    \label{Fig: coalitional_structure}
\end{figure*}

Note that even though being identical, the flexible units offer different individual flexibility areas. This is caused by locational effects, such as power losses and voltage constraints. For example, units C and D are located at the ends of feeders with low voltage profiles (buses 33 and 18). These units cannot increase their power consumption significantly due to the voltage limits, which results in reduced flexibility areas of coalitions \{C\} and \{D\}. It can also be seen how these constraints propagate in the coalitional structure once more players join the coalitions with units C and D.

The presented coalitional structure illustrates the maximum P-Q capacities of the units and the impact of network constraints. However, the flexibility areas estimated by model \eqref{Model boundary: objective}-\eqref{Model boundary: epsilon q} do not provide additional information on the operating conditions within the area, e.g., what units have to be activated, what level of coordination is required to reach certain operating points, and how to remunerate units for providing different flexibility requests. In this regard, in the rest of this section, the cost-minimising OPF model and the game-theoretic model are implemented to estimate the contributions of flexible units to specific flexible power requests.

\subsection{Allocation of Flexible Power Requests: a Cost-minimising OPF Approach}
\begin{figure*}[t]
    \centering
    \includegraphics[width=\textwidth]{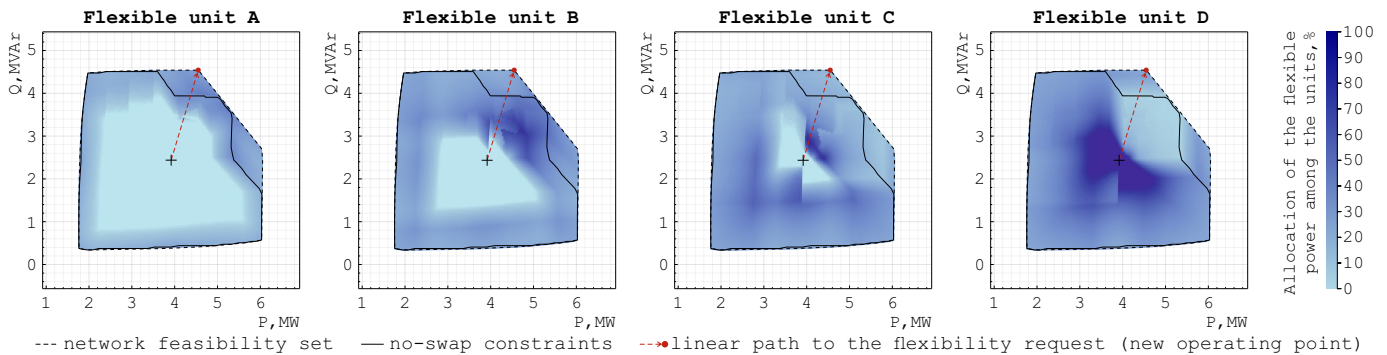}
    \caption{Allocation of the total apparent flexible power among the units for different flexibility requests according to the cost-minimising OPF model, in \%.
    }
    \label{Fig: cost_based_OPFs_multiple_flex_requests}
\end{figure*}

The allocation of flexible power among the units can be explicitly derived from the cost-minimising OPF model \eqref{Model cost: objective}-\eqref{Model cost: Q request} solved for any feasible flexible power request. The model indicates which units provide flexible power to meet the request while trying to activate the cheapest units first. Such allocations can differ significantly depending on the requested operating point in the P-Q space. Therefore, model \eqref{Model cost: objective}-\eqref{Model cost: Q request} was solved 14,520 times for different feasible flexibility requests (the feasible P-Q space was discretised by a grid with step 0.03 MVA). The resulting allocations are displayed in Fig.~\ref{Fig: cost_based_OPFs_multiple_flex_requests} as a percentage of the total apparent flexible power provided (in MVA). Note that many low-magnitude flexible power requests (close to the initial operating point) can be fully covered by unit D, which provides flexible power at the lowest cost. Therefore, for such requests, unit D is allocated up to 100\% of the requested power and should be paid much more than the other units.\footnote{Note that unit D is allocated less power than other units for the flexibility requests that increase the active and reactive power consumption of the network (the upper right side of the area). These differences stem from the voltage limitation at bus 18, which reduces the P-Q capability of unit D.} On the contrary, unit A has the highest cost and is not activated for many low-magnitude flexible power requests. It is activated and paid only for requests close to the flexibility area limits. Thus, under the cost-minimising OPF approach, the cheapest units get activated more often and are allocated more power and payments.


It can be observed from the cost-minimising allocation results that flexible power provision from several units has a highly nonlinear behaviour. There exist multiple shifts in the flexible power output of the units. These shifts occur due to both technical and economic reasons amplified by the nonlinearity of the network power flow model. For example, due to power losses and reactive power management, providing flexible power only by the cheapest units may not be the optimal solution for some operating points. For such points, model \eqref{Model cost: objective}-\eqref{Model cost: Q request} shifts a share of flexible power between the units to provide the least-cost feasible solution. Moreover, the analysis of flexible power allocations reveals the power swap phenomenon that happens when multiple units provide flexible power under network constraints. Specifically, some units can be producing flexible power, whereas other units consume flexible power. Such power swaps enable alleviating network constraints and reaching the limits of the network flexibility area. For example, in the 33-bus test system, a significant increase in the network power consumption cannot be achieved by increasing the power consumption of all flexible units since units C and D already operate close to the lower voltage limit of 0.9 p.u. Therefore, to reach the operating points close to the flexibility area boundary, model \eqref{Model cost: objective}-\eqref{Model cost: Q request} provides solutions where units C and D produce power. This enables alleviating the voltage limits at buses 33 and 18, whereas the remaining units consume flexible power to increase the network power consumption and meet the flexibility request.

The power swap phenomenon has not been analysed in the existing literature on the flexibility of distribution networks. Most of the studies imply perfect coordination of flexible units and focus on estimating the limits of the aggregated network flexibility, without specifying what actions are needed to reach the boundaries. However, the assumption of perfect unit coordination may not be realistic and requires further research, e.g., rapid shifts in flexible power allocation between close operating points impose additional ramp constraints on the unit operation. Using power swaps between flexible units to provide certain flexibility requests can also be controversial:\begin{itemize}
  \item From the operating standpoint, it can be more complex and less reliable to control flexible units working in different directions (producing and consuming power). For example, a failure of unit C or D to produce enough power while the overall network consumption is increased can result in voltage collapse of the distribution network.
  \item From the economic standpoint, power swaps lead to inconsistent solutions and issues with flexible units remuneration. For example, units can use much more flexible power due to simultaneous consumption and production than the total flexible power requested from DSO.
\end{itemize}

It is possible to forbid the power swap between flexible units by adding constraints \eqref{Power swap constraint: p}-\eqref{Power swap constraint: q} to the OPF model \eqref{Model cost: objective}-\eqref{Model cost: Q request}, forcing all units to either produce or consume power. The flexibility area with no flexible power swap is indicated in Fig.~\ref{Fig: cost_based_OPFs_multiple_flex_requests} by the solid line, while the entire flexibility area is denoted by the dashed line. It appears that network flexibility can be overestimated if not considering the issues related to flexible power swap. Moreover, imposing the no-swap constraints makes the aggregated flexibility areas nonconvex. Such nonconvexity could cause additional problems with distribution networks operation. For instance, a path between sequential operating points can contain infeasible points that require swapping power between flexible units.

\begin{figure}
    \centering
    \subfigure(a){\includegraphics[width=0.85\columnwidth]{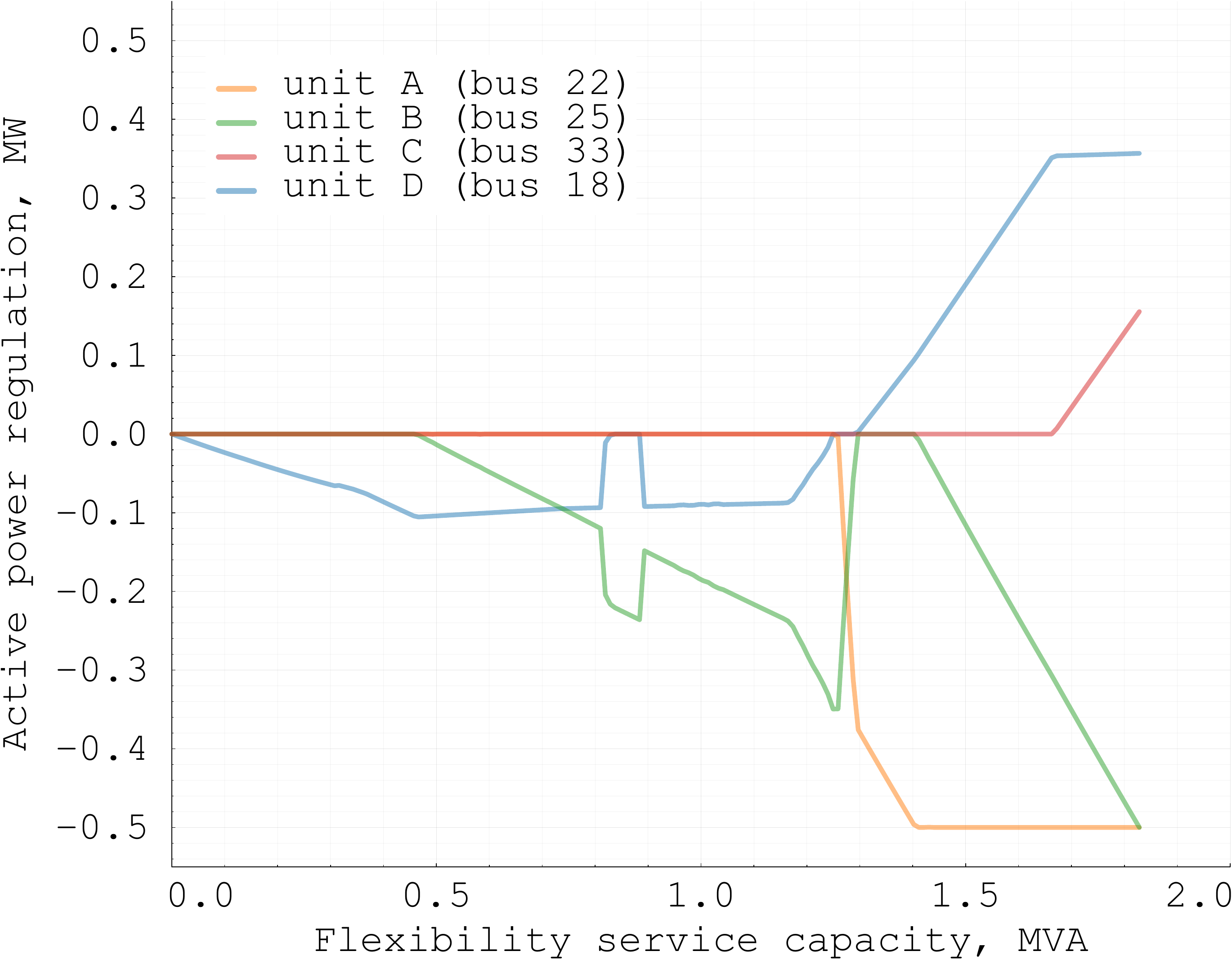}}
    \subfigure(b){\includegraphics[width=0.85\columnwidth]{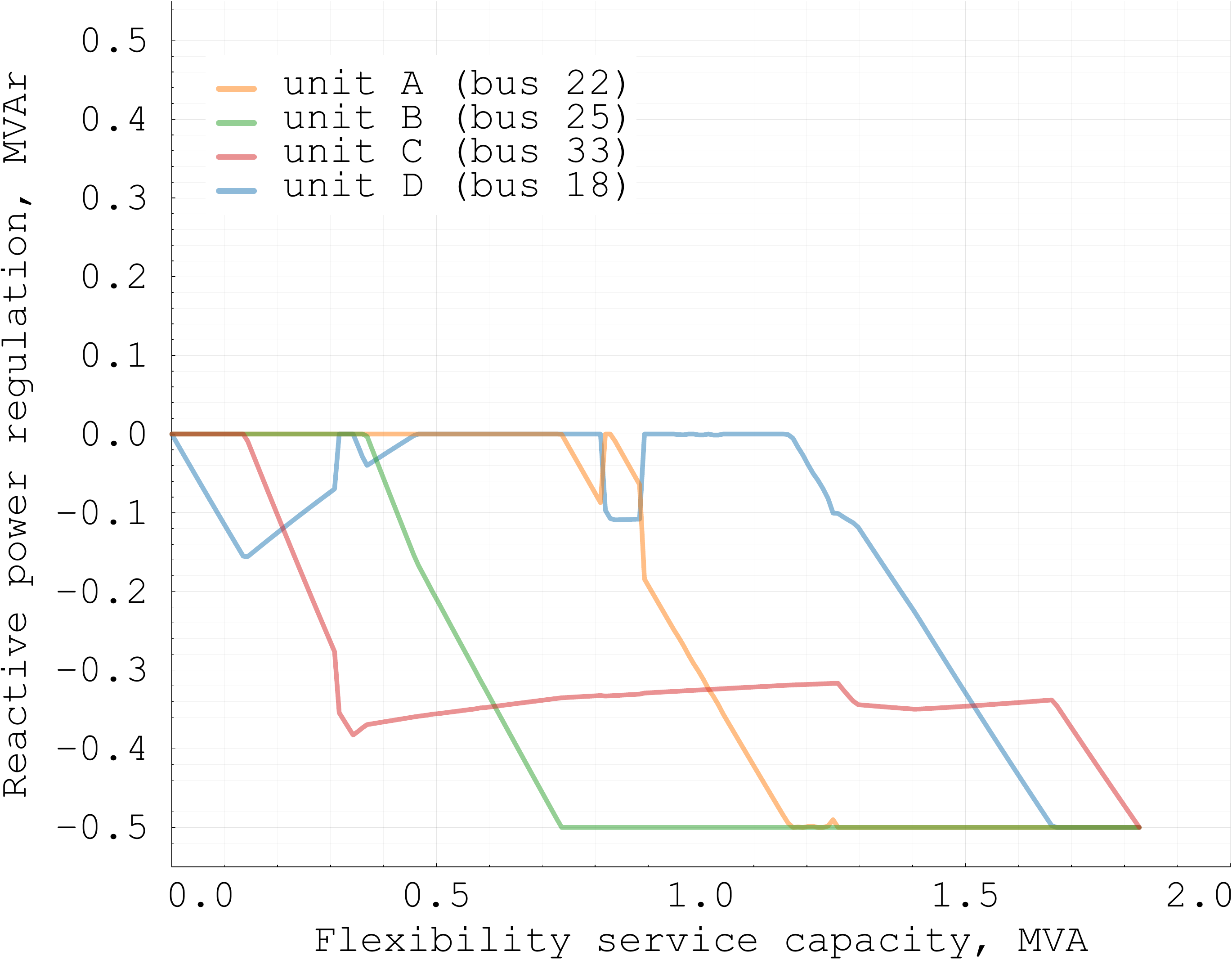}}
    \subfigure(c){\includegraphics[width=0.85\columnwidth]{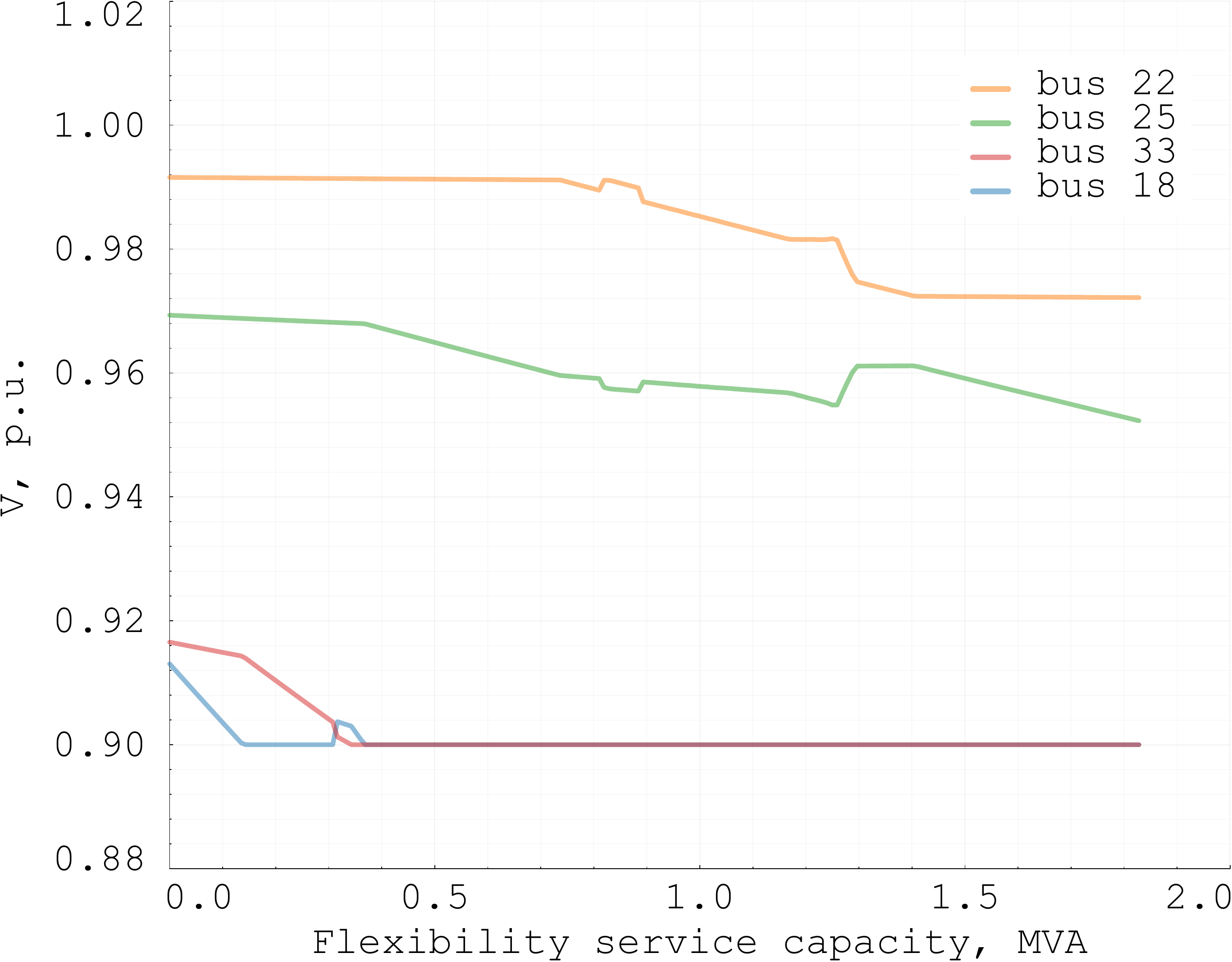}}
    \caption{Analysis of the optimal flexible unit dispatch along the linear flexibility service path defined in Fig.~\ref{Fig: cost_based_OPFs_multiple_flex_requests}: (a) flexible active power regulation, MW; (b) reactive power regulation, MVAr; (c) voltages at the ends of the feeders, p.u.}
    \label{Fig: path analysis}
\end{figure}

To further explore the flexible unit dispatch and coordination issues in flexibility provision, the simulations were performed for a series of operating points. A flexibility request (new operating point) was selected at the boundary of the flexibility area, as shown by the red dot in Fig.~\ref{Fig: cost_based_OPFs_multiple_flex_requests}. Then, the linear path from the initial operating point to the flexibility request was discretised with 200 steps. In each step, the cost-minimising model \eqref{Model cost: objective}-\eqref{Model cost: Q request} was solved to determine the optimal flexible power regulations by the units and voltage levels of the feeders. The results are displayed in Fig.~\ref{Fig: path analysis} as functions of the flexibility service capacity (the length of the path towards the selected operating point).

Analysis of the optimal flexible unit dispatch along the path shows that units have many permutations and rapid shifts in their P-Q regulations. Units C and D (the cheapest ones) are activated first to provide the flexibility service. But, voltages at buses 18 and 33 drop to the 0.9 p.u. limit, becoming the binding constraints and limiting further flexible power consumption by units C and D. For high-capacity flexibility services, the power swap effect can be observed: units C and D start producing flexible power while units A and B consume it. Other nonlinear changes and rapid shifts in the flexible P-Q regulations are associated with the capacity limits of the units and differences in their cost functions. For instance, at a flexibility service capacity of 0.72 p.u., unit B reaches its maximum reactive power regulation, which causes unit A to start consuming power. The same effect happens when unit A reaches its reactive power regulation limit at flexibility service capacity of 1.15 p.u. At service capacities of 0.8 and 0.9 p.u. there are P-Q regulation shifts indicating economic trade-offs between consuming active and reactive flexible power.

\begin{figure}
    \centering
    \includegraphics[width=\columnwidth]{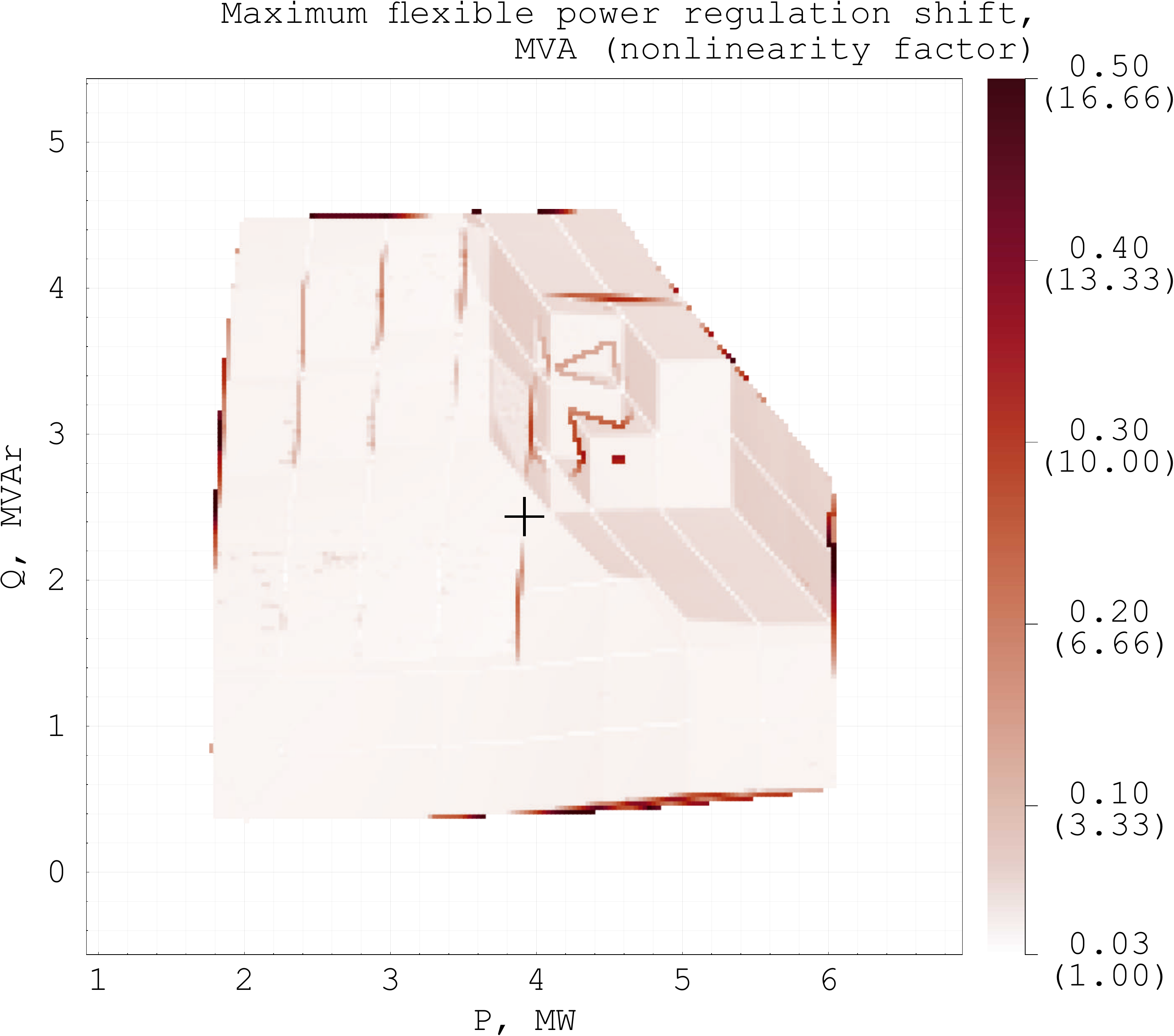}
    \caption{Nonlinearity assessment of the optimal unit dispatch. The red heatmap displays areas with high nonlinearity factors where at least one unit shifts its flexible power regulation rapidly.}
    \label{Fig: nonlinearity_segments_analysis}
\end{figure}

The observed nonlinear and rapid shifts in the flexibility provision pose challenges for the DSO operation. To identify and classify these shifts systematically, the nonlinearity metric introduced in \eqref{Nonlinearity max} and \eqref{Nonlinearity factor} was applied to the network P-Q feasibility space discretised with step 0.03 MVA. The nonlinearity assessment is shown in Fig.~\ref{Fig: nonlinearity_segments_analysis}, where the dark red regions indicate the edges of rapid nonlinear shifts in the flexible power regulation. In some cases, flexible units change their power by 0.5 MVA, which is 16.66 times more than the distance between neighbouring operating points analysed. The DSO can avoid these regions with high nonlinearity factors to ensure the security and economic efficiency of flexibility services provision. 
Note that the performed nonlinearity assessment is similar to the edge detection via kernel (convolution matrix) used in image processing and computer vision \cite{Dai2017,Krizhevsky2017}. It, therefore, offers opportunities for developing Machine Learning models to analyse the effects of nonlinear constraints and predicting undesirable operating points with rapid changes in the flexible unit dispatch.

This subsection demonstrated the usefulness of the cost-minimising models for flexibility tracing and valuation. Such models enable finding the least-cost flexible unit dispatch and identifying flexible power swaps between units and rapid changes in flexible power regulation. However, a purely cost-based analysis of network flexibility has several disadvantages. First, cost-minimising models consider only a single outcome of flexibility provision for each operating point. Other possible combinations of units and their contributions are neglected. Therefore, such models cannot be used to comprehensively rank units and assess their criticality. Second, remuneration mechanisms based on cost-minimising models always favour the cheapest units and do not give incentives for more expensive units to participate in the flexibility market. In this regard, the next subsection illustrates the advantages of the proposed game-theoretic approach, which captures possible contributions of units and enables including additional metrics of flexibility provision.

\subsection{Allocation of Flexible Power Requests: a Game-Theoretic Approach}
Any feasible flexible power request can be analysed using the cooperative game framework, where the grand coalition fully provides the required flexible power, and the subcoalitions maximise their flexible power provision in the required direction in the P-Q space. As discussed in Section \ref{Subsection: flexibility as a cooperative game}, different metrics can be used to characterise coalitions, such as the flexible power provided, its cost, or its economic surplus. Then, the Shapley value can be applied to estimate the contributions of units and perform their ranking and valuation. This subsection presents simulations for multiple flexible power requests based on the game-theoretic approach and analyses the allocation of flexibility using different metrics.

\begin{figure*}
    \centering
    \includegraphics[width=\textwidth]{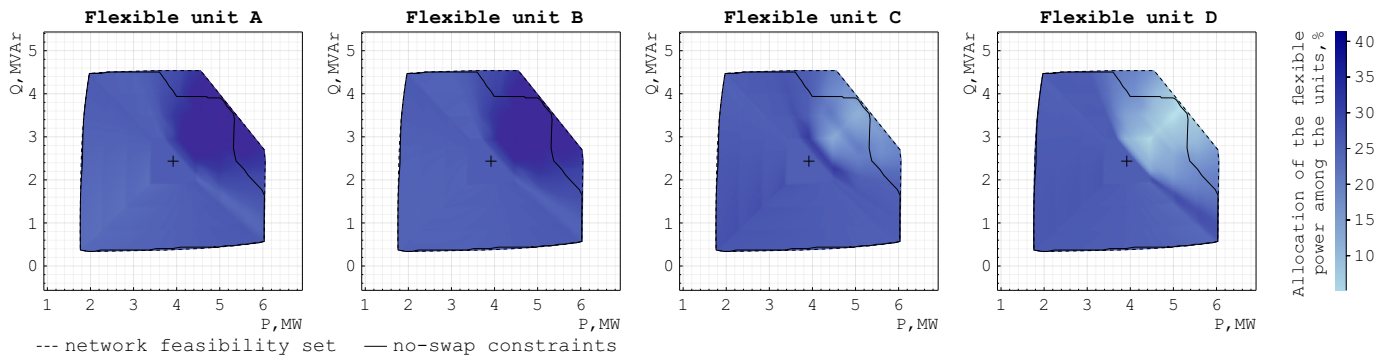}
    \caption{Allocation of the total apparent flexible power among the units for different flexibility requests according to the Shapley value, in \%.
    }
    \label{Fig: Shapley_values_multiple_flex_requests}
\end{figure*}

First, the aggregated apparent flexible power (in MVA) is selected as the metric to characterise coalitions, i.e., each coalition is represented by the P-Q limits that flexible units can reach for a given operating point and power factor. This capacity-based metric enables ranking units by their contributions to flexibility requests and identifying the most critical units.\footnote{In this context, the criticality of a flexible unit can be seen as the priority ranking of its contributions to a given flexibility request. More critical units contribute the most to flexible power provision and therefore have the greatest impact on the flexibility service reliability. Inaccurate forecasting or unavailability of such units can result in the infeasibility of the related operating points.} The cooperative game is solved using the Shapley value for multiple feasible flexibility requests, as displayed in Fig.~\ref{Fig: Shapley_values_multiple_flex_requests}. The results indicate that for low-magnitude requests, the units are equally useful and get allocated 25\% of the requested flexible power. This happens because the units can provide the same flexible power for such requests and bring equal contributions to the coalitions. Thus, they are symmetric players in the cooperative game. However, a significant divergence in the contributions happens for flexibility requests that increase the power consumption of the network. For such requests, units C and D cannot contribute much since their ability to increase power consumption is limited due to the voltage constraints. These limitations are captured by the cooperative game formulation and result in the lower ranking for units C and D. Note that such differences stem from the individual P-Q capabilities of the units, as previously illustrated by the coalitional structure in Fig.~\ref{Fig: coalitional_structure}.

\begin{figure*}
    \centering
    \includegraphics[width=\textwidth]{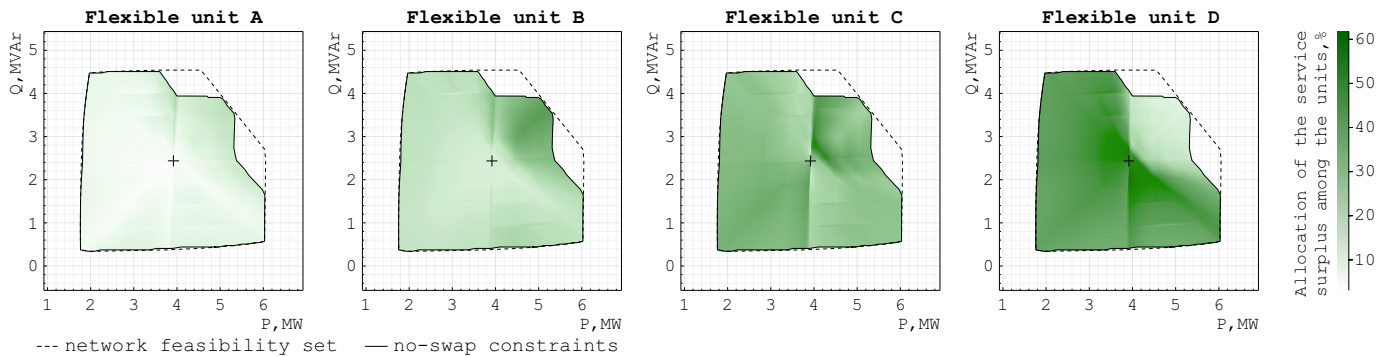}
    \caption{Allocation of the flexibility service economic surplus among the units for different flexibility requests according to the Shapley value, in \%.
    }
    \label{Fig: surplus_based_cooperative_games_multiple_flex_requests_NO_SWAP}
\end{figure*}


Second, the economic surplus of providing flexible power (in \$/h) is selected to characterise coalitions. This is a complex metric that captures both the P-Q capabilities of the units and their economic impact, i.e., the potential profits that they bring when providing flexible power. The resulting economic surplus allocations for multiple flexibility requests obtained by the Shapley value are displayed in Fig.~\ref{Fig: surplus_based_cooperative_games_multiple_flex_requests_NO_SWAP}. These simulations reflect the cost causality in flexibility valuation and indicate which units contribute to the surplus of the flexible power provision.\footnote{Note that only the flexibility area with no power swap is considered to test this approach. Power swaps between flexible units can lead to inconsistent solutions, where units simultaneously produce and consume much more flexible power than requested from an ADN. Such solutions can be significantly more costly than the nearby operating points with no power swap, even making the surplus of the flexible power provision negative. Analysis of the flexible units cooperation under power swaps is left for future research.} For instance, unit D can produce flexible power at the lowest cost and contributes the most to the surplus of the flexible power provision, except in cases of the network consumption increase, where unit D cannot provide much flexible power.




The allocation solutions estimated by the game-theoretic approach for different metrics provide a variety of options to rank flexible units and remunerate them. The capacity-based allocation estimates the criticality of units for a specific flexibility request. The surplus-based game-theoretic approach explicitly considers the cost of flexible power provision and estimates the economic impact of flexible units. This approach can serve as the remuneration mechanism: it allocates more economic surplus to units that provide more flexible power at lower costs and get activated in many possible coalitions. Note that unlike the allocations obtained by the cost-minimising OPF model, the solutions provided by the game-theoretic approach have no operating points where one of the units is allocated 100\% of the cooperation value. This happens since the cooperative game formulation considers not only the flexible power of units in the grand coalition but also their contributions to all possible subcoalitions. 
The advantages and the applicability of the proposed game-theoretic framework are further discussed in Section~\ref{Section: discussion}.

\section{Discussion}\label{Section: discussion}
\subsection{Potential Applications}
The above simulations demonstrate how the proposed framework can be used to meet various needs of DSOs. The capacity-based analysis of network aggregated flexibility enables DSOs to identify the most critical units for certain flexibility requests. The cost-based and surplus-based approaches incorporate more complex metrics and capture both the outputs of flexible units activated and their contributions to the value of cooperation. Such approaches can be used not only for ranking the usefulness of units but also for flexibility remuneration mechanisms. The cost-minimising OPF model \eqref{Model cost: objective}-\eqref{Model cost: Q request} offers straightforward solutions where the cheapest units are activated and paid first. Despite its simplicity, this model does not account for possible contributions of more expensive units and does not consider scenarios where the cheapest units may not be available. 
In this regard, the allocation of the flexibility service economic surplus among flexible units according to the Shapley value can constitute a more promising remuneration mechanism.

\begin{figure*}
    \centering
    \includegraphics[width=\textwidth]{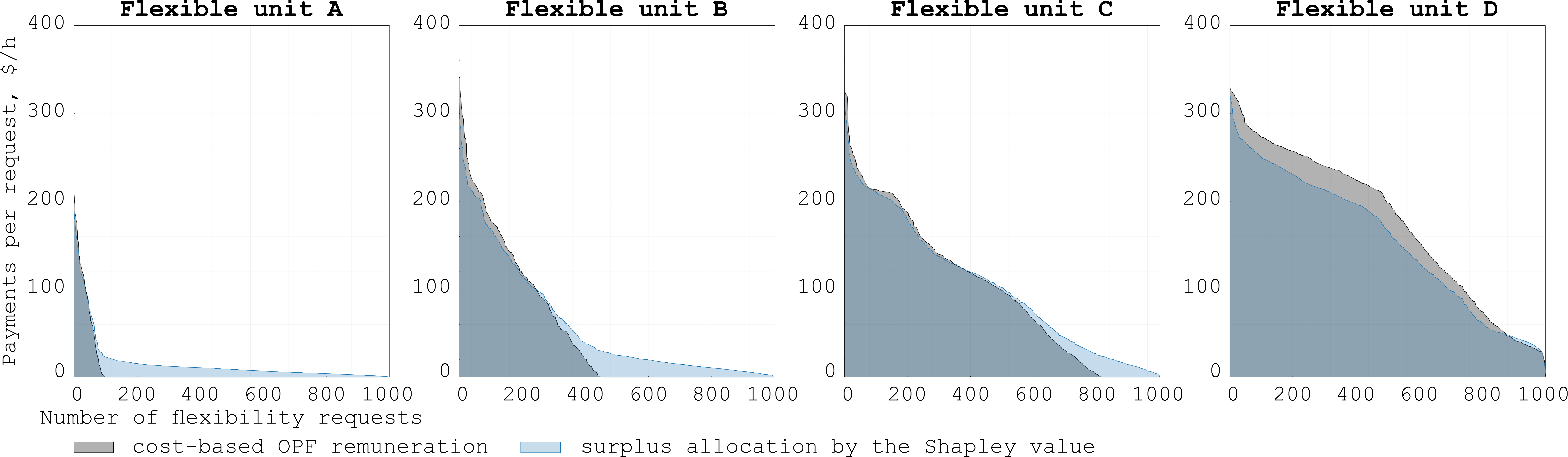}
    \caption{Payments to flexible units for 1,000 flexibility requests under different remuneration schemes. The payments are arranged in descending order for each unit. The grey plots correspond to the remuneration scheme based on the cost-minimising OPF model, where the cheapest units are activated and paid first. The blue plots present the remuneration scheme with the allocation of flexibility service economic surplus according to the Shapley value.}
    \label{Fig: payment_histograms}
\end{figure*}

To support the Shapley-based remuneration mechanism, Fig.~\ref{Fig: payment_histograms} presents the comparison of the payments allocated to flexible units with the payments obtained from the cost-minimising OPF model. A thousand randomly generated flexibility requests were simulated. A normal probability distribution was assumed with the mean set at the initial operating point of the network and the standard deviation of 0.6 MVA. The simulations reflect realistic flexibility requests, where low-magnitude flexible power deviations are requested more often than high-magnitude ones. According to the cost-minimising OPF model, the cheapest unit D is activated and paid for each of the simulated flexibility requests. Other units provide flexibility less frequently and receive far fewer payments. Unit A, the most expensive one, is called in only about 1/10 of all flexibility requests. Such payments might not incentives unit A to participate in the flexibility market. However, this unit is still valuable for the provision of flexibility. 
Its contributions are considered in the Shapley-based remuneration mechanism, which always allocates to unit A a share of the flexibility service economic surplus. In this way, more expensive units get additional incentives to participate in the market and declare their real P-Q capabilities and costs. Some recent studies, e.g., \cite{Ruwaida2022,Sarstedt2022}, highlighted the need to introduce availability remuneration in flexibility markets. In this regard, the Shapley value, which considers both the optimal unit dispatch and units' potential contributions in different coalitions, can serve as a mechanism for determining the availability and delivery (utilisation) payments.

\subsection{Incentive Compatibility in Aggregated Flexibility Pricing}
To ensure efficient procurement and optimal utilisation of flexibility services, it is important for DSOs to implement flexibility pricing mechanisms with the incentive compatibility property \cite{Tsaousoglou2021}. Incentive-compatible mechanisms guarantee that for each flexible unit, truthful reporting of information (such as flexible power cost and the available capacity) becomes the profit-maximising dominant strategy. Thus, flexibility market participants do not have incentives to misreport their true data or deviate from the declared values.

To analyse the incentive compatibility of the Shapley value as a pricing mechanism for aggregated flexibility services, it is necessary to define individual and coalitional rationality for players (these constraints are known as the Core of the game), verify the non-emptiness of the Core and the convexity of the game, compute the excess of coalitions and other incentives for players to join them \cite{Anibal2022,Churkin2022enhancing}.\footnote{{A cooperative game with transferable utility is called convex if the marginal contribution of any fixed player $i$ to coalition $S$ increases as more players join the coalition. Convexity is the desired property of cooperative games that guarantees the nonemptiness of the Core. It also guarantees that the Shapley value is an element of the Core (i.e., is individually and coalitionally rational for all players).}}
Note that in the proposed framework, cooperative games change for each feasible flexibility request, involving different network constraints and flexibility market outcomes.
There exist operating points that lead to cooperative games with the Shapley value providing not incentive-compatible payment allocations. For example, in operating points that require flexible power swaps between units, some subcoalitions have high costs due to economically not effective power swaps, which makes the cooperative game non-convex and the Shapley value allocation not incentive-compatible. It is important to find conditions under which the Shapley value provides incentive-compatible pricing for aggregated network flexibility. A thorough analysis of the Shapley-based flexibility pricing, its incentive compatibility and manipulability is the subject of future research.

\subsection{Scalability and Applicability Issues}
Cooperative game formulations have the disadvantage of being prone to scalability issues. This is because the number of possible coalitions in a cooperative game, $2^N$, increases exponentially with the number of players $N$ (flexible units available in a network). To describe each coalition, model \eqref{Model game: objective}-\eqref{Model game: binaries-coalitions} has to be solved, and the selected metric should be derived. Therefore, for games with hundreds or thousands of players, it can be intractable to consider all coalitions and implement the Shapley value formula \eqref{CGT: Shapley} directly. 
{\color{black} The challenges of the Shapley value scalability have been recognised by literature in Game Theory, power systems research, and data science (as the Shapley value has been found useful for interpreting machine learning models \cite{Jia2020,Mitchell2022,Okhrati2020}). As a result, several approaches have been proposed to estimate the Shapley value (with different levels of accuracy) while minimising calculations to facilitate its scalability. For example, the Shapley value estimations based on random sampling (or stratified random sampling) \cite{Castro2009,Castro2017,Maleki2013,agarwal2019marketplace,Goncalves2021,Cremers2023} reduce computational costs by considering a limited number of coalitions instead of explicitly modelling $2^N$ coalitions. However, the drawback of random sampling is its slow convergence. For games with thousands of players, random sampling may require an unreasonably high number of simulations to accurately approximate the Shapley value. In this regard, advanced sampling and permutation methods have been proposed to improve the convergence of the Shapley value approximations \cite{Jia2020,Azuatalam2019,Mitchell2022,Okhrati2020}. For example, compressed sensing \cite{Jia2020} can exploit the sparse structure of the player's contributions in cooperative games, and quadrature methods and kernel functions \cite{Mitchell2022} enable generating targeted samples with good properties.
Also, the Shapley value can be applied to large problems by reducing the number of players in cooperative games via clustering algorithms \cite{Azuatalam2019,Cremers2023}. That is, clusters of players can be considered instead of modelling individual players. But accuracy of such applications is highly dependent on the choice of clustering algorithm.
Finally, the Shapley value can be estimated by using alternative representations, extensions, or decompositions of cooperative games \cite{Owen1972,Ieong2005,STERN2019,Okhrati2020}.
It follows that, with the recent advances in the Shapley value estimation approaches, it can be applied to solve practical cooperative games with hundreds or thousands of players.
}

{\color{black}In the context of aggregated DER flexibility valuation,} the practical limit of the exact enumeration of possible coalitions and implementation of the Shapley value formula \eqref{CGT: Shapley} is roughly 10 flexible units, which requires solving 1023 OPF problems. Considering the performance of modern algorithms and solvers, each OPF can be solved in seconds or even fractions of a second for an average distribution network. Thus, with a single standard computer, the proposed flexibility valuation mechanism can take up to 20-30 minutes to allocate payments for flexible units. Moreover, since the OPF problems for different coalitions of units are independent, the allocation process can be sped up using parallel computing. Therefore, applying the proposed mechanism to flexibility remuneration in intraday 30-minute and hourly markets should be realistic.

{\color{black}For larger numbers of units, e.g., 1000 independent flexibility providers, the proposed methodology could be extended with the aforementioned random sampling and clustering techniques to approximate the Shapley value by solving only a few thousand OPF problems. First, 1000 flexibility providers can be clustered based on their location and parameters into a few dozen players. Then, the cooperative game with dozens of players can be solved using the Shapley value approximations, e.g., based on advanced sampling and permutation methods. Finally, the value of cooperation allocated to each player by the Shapley value approximation should be further divided among the flexible units in the clusters. A similar approach has already been applied in studies on energy communities. In \cite{Cremers2023}, clustering of prosumers within energy communities was explored, and the Shapley value approximations were used to allocate benefits among them. The simulations presented in \cite{Cremers2023} demonstrate that this approach can be applied to cases with up to 200 consumers, with an approximation error below 1\%. 
Therefore, the Shapley-based valuation approach proposed in this work can be applied to cases with hundreds of flexible units. However, further research is needed to verify the computational efficiency and accuracy of the clustering and approximation methods in the context of aggregated DER flexibility valuation.
}

\section{Conclusion}\label{Section: conclusion}
This work investigates the formation of network aggregated flexibility and proposes a framework for tracing, ranking, and valuation of aggregated flexible power within ADNs. 
A set of models is introduced for estimating the limits of aggregated network flexibility provision in the P-Q space, minimising the cost of flexibility services, and capturing contributions of flexible units to aggregated flexibility via the cooperative game formulation.
Extensive simulations performed for numerous feasible operating points of the 33-bus radial test system demonstrate the effectiveness of the proposed framework for aggregated flexibility ranking and valuation. The simulations also illustrate the principles of aggregated flexibility formation, the effects of network constraints, and the nonlinear complex behaviour of flexible units.
The flexible power swap effect is discovered, which happens when different units simultaneously produce and consume flexible power to alleviate network constraints and maximise the flexibility service provided by a distribution network.
The nonlinearity metric and the no-swap constraints are implemented to analyse the optimal unit dispatch and potential coordination issues.

The proposed framework incorporates different metrics of flexibility and can be used by DSOs in the following applications. First, ranking flexible units by their contributions to the aggregated network flexibility identifies the most critical units in the network and provides information on the structure and diversification of flexible resources. Second, cost-based and surplus-based ranking can serve as a remuneration mechanism for flexible units. As discussed in the paper, the surplus-based allocation mechanism can give flexible units incentives to declare their maximum capability at a lower cost. However, the combinatorial nature of flexible power aggregation makes the cooperative game formulation intractable for cases with hundreds or thousands of flexible units. Future research will try to overcome these limitations by using advanced clustering and compression techniques, Shapley value approximations, and cooperative game decompositions. It is also crucial to investigate the incentive compatibility and manipulability of the Shapley-based flexibility pricing.

\begin{figure}
    \centering
    \includegraphics[width=0.9\columnwidth]{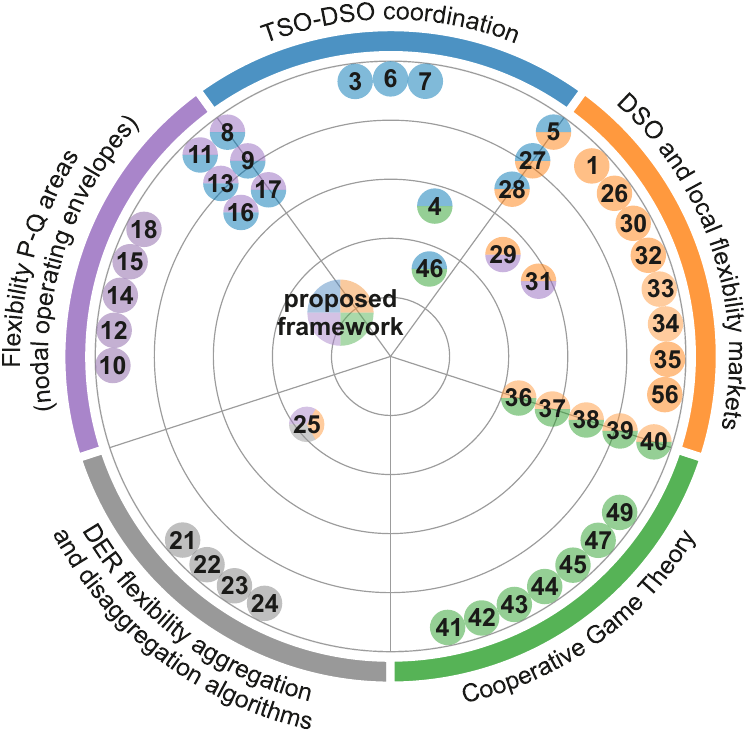}
    \caption{Mapping of the most relevant references and research directions.}
    \label{Fig: literature_review_radar}
\end{figure}

\appendix[Mapping of Literature and Research Directions]\label{Appendix}
To illustrate the research gap in the existing literature and highlight the contributions of this work, Fig.~\ref{Fig: literature_review_radar} presents a mapping of the most relevant references and research directions. The figure defines five research directions and arranges references according to their contributions to these topics.
It follows that the proposed framework includes Cooperative Game Theory, TSO-DSO coordination, and flexibility markets, with a stronger focus on ADN operation and aggregated P-Q flexibility.
By combining these areas, the proposed framework enables estimating contributions of flexible units to aggregated network flexibility and to each separate flexibility request, ranking the criticality of flexible units for the flexibility service provision, analysing the nonlinearity of the optimal flexible unit dispatch and potential unit coordination issues, and introducing flexibility remuneration mechanisms that include availability and delivery (utilisation) payments.
The closest studies \cite{Vicente-Pastor2019,Anibal2022} did not use the concept of flexibility P-Q areas to analyse flexible power aggregation within distribution networks, and study \cite{Sarstedt2022} did not use Cooperative Game Theory to propose tracing, ranking and valuation mechanisms for aggregated network flexibility.


%





\ifCLASSOPTIONcaptionsoff
  \newpage
\fi



%




\bibliographystyle{IEEEtran}
\bibliography{references.bib}

%








\end{document}

%% file: packages.tex
\usepackage{amsmath, amssymb, amsfonts, amsthm, booktabs, cancel, comment, enumitem, eurosym, float, graphicx, mathtools, multirow, nicefrac, placeins, soul, steinmetz, stfloats, siunitx, svg, url, xcolor, tabularx}%
\usepackage{commath}
\usepackage{cite}
\usepackage{subfigure}
\usepackage{hyperref}
\usepackage{nomencl}
\usepackage{cleveref}
\makenomenclature

\floatstyle{ruled}
\newfloat{model}{thp}{lop}
\floatname{model}{\textsc{Model}}

\makeatletter
\renewcommand*{\fnum@model}{\fname@model}
\makeatother